\documentclass[12pt,aip,pas,showpacs,reprint]{revtex4-1}
\usepackage{CJK}
\usepackage{graphicx}
\usepackage{todonotes}
\usepackage{geometry}
\usepackage{mathrsfs}
\usepackage{subfigure}
\usepackage{amssymb}
\usepackage{indentfirst}
\usepackage{color}
\usepackage{cancel}
\usepackage{amsmath}
\usepackage{hyperref}
\usepackage{float}
\begin{document}
\begin{CJK*}{GB}{}
\title{\LARGE\textbf{Random Scalar Fields and Hyperuniformity}}
\author{Zheng Ma}
\affiliation{Department of Physics, Princeton University, Princeton, New Jersey 08544, USA}
\author{Salvatore Torquato}
\email{torquato@electron.princeton.edu} 
\affiliation{Department of Physics, Princeton University, Princeton, New Jersey 08544, USA}
\affiliation{Department of Chemistry, Princeton University, Princeton, New Jersey 08544, USA}
\affiliation{Princeton Institute for the Science and Technology of Materials, Princeton University\\Princeton, New Jersey 08544, USA}
\affiliation{Program in Applied and Computational Mathematics, Princeton University\\Princeton, New Jersey 08544, USA}
\begin{abstract}
\noindent Disordered many-particle hyperuniform systems are exotic amorphous states of matter that lie between crystals and liquids. Hyperuniform systems have attracted recent attention because they are endowed with novel transport and optical properties. Recently, the hyperuniformity concept has been generalized to characterize two-phase media, scalar fields and random vector fields. In this paper, we devise methods to explicitly construct hyperuniform scalar fields. 
Specifically, we analyze spatial patterns generated from Gaussian random fields, which have been used to model the microwave background radiation and heterogeneous materials, the Cahn-Hilliard equation for spinodal decomposition, and Swift-Hohenberg equations that have been used to model emergent pattern formation, including Rayleigh-B{\' e}nard convection. We show that the Gaussian random scalar fields can be constructed to be hyperuniform. We also numerically study the time evolution of spinodal decomposition patterns and demonstrate that they are hyperuniform in the scaling regime. Moreover, we find that labyrinth-like patterns generated by the Swift-Hohenberg equation are effectively hyperuniform. We show that thresholding (level-cutting) a hyperuniform Gaussian random field to produce a two-phase random medium tends to destroy the hyperuniformity of the progenitor scalar field. We then propose guidelines to achieve effectively hyperuniform two-phase media derived from thresholded non-Gaussian fields. Our investigation paves the way for new research directions to characterize the large-structure spatial patterns that arise in physics, chemistry, biology and ecology. Moreover, our theoretical results are expected to guide experimentalists to synthesize new classes of hyperuniform materials with novel physical properties via coarsening processes and using state-of-the-art techniques, such as stereolithography and 3D printing.
    
\end{abstract}
\date{}
\maketitle
\end{CJK*}
\newpage

\section{INTRODUCTION}

\indent Disordered many-particle hyperuniform systems are exotic amorphous states of matter that lie between crystals and liquids. They are like liquids in that they are statistically isotropic with no Bragg peaks, but they also resemble perfect crystals in the way they suppress large-scale density fluctuations. The theoretical foundations of hyperuniformity was only introduced around a decade ago, \cite{torquato2003local} and a comprehensive understanding of hyperuniformity is still in its infancy. The basic feature of a many-particle hyperuniform system is the suppression of density fluctuations at large length scales, which provides a new way to classify crystals, quasicrystals and special disordered systems. \cite{torquato2003local, zachary2009hyperuniformity} In $d$-dimensional Euclidean space $\mathbb{R}^d$, we know that for a Poisson gas, the number variance $\sigma_N^2(R)$ of particles within a spherical observation window of radius $R$ has the volume scaling: $\sigma_N^2(R)\sim R^d$, which is also true for most typical disordered systems such as liquids. A translationally invariant (statistically homogeneous) hyperuniform point configuration is one in which the number variance grows more slowly than $R^d$ or, equivalently, it possesses a structure factor that satisfies the following condition:
\begin{equation} \label{struc}
\lim_{|\mathbf k|\rightarrow 0} S(\mathbf k)=0.
\end{equation} 
Perfect crystals and many quasicrystals are hyperuniform with the surface-area scaling $\sigma_N^2(R) \sim R^{d-1}$. There are disordered hyperuniform systems that have the same scaling as crystals but without translational or rotational symmetries. If the structure factor goes to zero with the power-law form as the wavenumber $k\equiv |{\mathbf k}|$ tends to zero, i.e.,
\begin{equation} \label{strucscal}
S(\mathbf k)\sim |\mathbf k|^{\alpha} \qquad |{\mathbf k}| \rightarrow 0,
\end{equation} 
where $\alpha>0$, the number variance has the following large-R asymptotic scaling: \cite{torquato2003local}
\begin{equation} \label{nv}
\sigma_N^2(R)\sim \left \{ \begin{aligned} &R^{d-1}, &\alpha>1,\\ &R^{d-1}\ln R, &\alpha=1,\\&R^{d-\alpha}, &0<\alpha<1.\end{aligned}\right.   (R\rightarrow \infty) 
\end{equation}
The exponent $\alpha$ is a rough measure of short-range order. As $\alpha$ increases from zero, the degree of short-range order generally increases. Indeed, in the limit $\alpha \rightarrow \infty$,
a hyperuniform system becomes ``stealthy", \cite{batten2008classical, torquato2015ensemble, zhang2015ground} which means there is no scattering in an exclusion region around the origin in ${\bf k}$-space, i.e.,
\begin{equation} \label{stealthy}
S(\mathbf k)=0  \qquad \mbox{for} \ 0\leq |\mathbf k|\leq K.
\end{equation}
Stealthy configurations include all perfect crystals, special disordered hyperuniform structures, and
so-called aperiodic ``stacked-slider" phases. \cite{zhang2015ground}\\
\indent Disordered hyperuniform systems have been found in many physical systems during the last decade. Remarkably, these examples span both equilibrium and nonequilibrium phases, including both classical and quantum varieties. Examples include ``stealthy" disordered ground states, \cite{torquato2015ensemble} maximally random jammed particle packings, \cite{PhysRevLett.95.090604, zachary2011hyperuniform, atkinson2016critical} jammed colloidal glasses, \cite{berthier2011suppressed, kurita2011incompressibility, dreyfus2015diagnosing} dynamical processes in ultracold atoms, \cite{PhysRevA.90.011603} disordered networks with large photonic band gaps, \cite{florescu2009designer} and driven nonequilibrium systems. \cite{PhysRevLett.114.110602} Moreover, hyperuniformity arises in biological systems and mathematics, such as avian photoreceptor patterns, \cite{PhysRevE.89.022721} immune system receptors \cite{mayer2015well} and even the one-dimensional point patterns derived from the nontrivial zeros of the Riemann zeta function. \cite{montgomery1973pair} It is expected that new discoveries will continue to emerge since we now know what to look for, namely, anomalous suppression of
large-scale density fluctuations.\\           
\indent The practical implications of hyperuniformity are also noteworthy. Hyperuniform materials have been fabricated at the micro- and nanoscales for various photonic applications, \cite{man2013isotropic, PhysRevA.88.043822, Man:13} surface-enhanced Raman spectroscopy, \cite{C4CP06024E} realization of a terahertz quantum cascade laser, \cite{doi:10.1117/12.2083678} self-assembly of diblock copolymers, \cite{PhysRevE.92.050601} and dense transparent materials. \cite{Leseur:16} Moreover, the degree to which a disordered system is hyperuniform provides a useful means of characterizing large-scale
structural order. \cite{hejna2013nearly, xu2016influence}\\
\indent The hyperuniformity notion has been recently generalized to heterogeneous media (e.g., composites and porous media), \cite{zachary2009hyperuniformity,0953-8984-28-41-414012} random scalar fields (e.g., temperature and concentration fields) \cite{PhysRevE.94.022122} and random vector fields (e.g., fluid velocities in porous media and turbulent flow). \cite{PhysRevE.94.022122} The latter category requires the extension of the standard definition of hyperuniformity to depend on the direction at which the origin is approached in Fourier space.\\ 
\indent Since the analysis of random scalar fields is more straightforward than that of random vector fields, and heterogeneous media can also be seen as special cases of random scalar fields, this paper focuses on the further development of our understanding of the hyperuniformity of scalar fields. Scalar fields can arise in a variety of physical contexts, including concentration and temperature fields in heterogeneous and porous media, \cite{torquato2013random} laser speckle patterns, \cite{cite-key} and spatial patterns in biological, chemical and ecology systems. \cite{PhysRevA.15.319, 59776, bonachela2015termite, liu2013phase} Torquato \cite{PhysRevE.94.022122} showed that scalar fields generated by convolving hyperuniform point configurations with a non-negative dimensionless radial scalar kernel function will inherit the hyperuniformity of the original point configurations.\\
\indent In this paper, we delve more deeply into the study of hyperuniform scalar fields, especially how to construct them via Gaussian random fields that have been used to model a variety of systems, including the microwave background radiation, \cite{liddle2000cosmological} heterogeneous materials, \cite{roberts1995transport} and random speckle fields, \cite{dainty1977statistics} Cahn-Hilliard equations for spinodal decomposition, \cite{rundman1967early} and Swift-Hohenberg equations that describe thermal convection in hydrodynamics \cite{PhysRevA.15.319} as well as a general model of emergent pattern formation. \cite{cross2009pattern} We re-examine these well-known models under the ``hyperuniform" lens and establish the conditions under which they are hyperuniform or not.\\
\indent We start by examining Gaussian random fields  and show how to construct hyperuniform fields from them. We then 
study whether time-evolving patterns that arise from spinodal decomposition via the Cahn-Hilliard description, \cite{cahn1958free} which are ubiquitous in chemistry \cite{rundman1967early} and biological systems, \cite{elson2010phase} are hyperuniform. We find that these patterns are actually hyperuniform in the scaling regime. Other nonequilibrium spatial patterns with well-defined characteristic wavelengths are also briefly mentioned. Torquato \cite{0953-8984-28-41-414012} suggested that some thresholded Turing patterns may not be hyperuniform, but this is not necessarily true for general scalar fields. Here we show that the Swift-Hohenberg equations can yield disordered hyperuniform patterns. We also present a toy ``polycrystal" model to describe the labyrinth-like patterns generated by the Swift-Hohenberg equation. Subsequently, we investigate whether a two-phase system that is created by thresholding a general hyperuniform scalar field can inherit the hyperuniformity of the parent system. We theoretically ascertain that a thresholded disordered Gaussian random field can never be hyperuniform. Then we propose possible ways to achieve hyperuniform two-phase media from thresholding and provide a numerical example. Our theoretical results can now be used to guide experimentalists to synthesize new classes of hyperuniform materials via coarsening processes directly or stereolithography \cite{melchels2010review} and/or 3D printing techniques. \cite{laurin2014hollow} There is a great need to expand experimental capabilities in order to make large samples of hyperuniform materials and our work expands those experimental horizons. The results of our study highlight the richness of the hyperuniformity concept and provides new paths to construct them theoretically and to synthesize them experimentally.\\
\indent In Sec. II, we provide necessary theoretical definitions and background. In Sec. III, we explicitly show how to generate hyperuniform scalar fields via Gaussian random fields, the Cahn-Hilliard equation for spinodal decomposition and Swift-Hohenberg equations. In Sec. IV, detailed simulation procedures for the aforementioned systems and their corresponding numerical results are provided. In Sec. V, we discuss how to obtain two-phase media via thresholding random scalar fields and whether they can be hyperuniform. We show that thresholding hyperuniform Gaussian fields generally cannot produce hyperuniform two-phase media. Finally, in Sec. VI, we make concluding remarks and discuss the implications of our findings.\\     

\section{BACKGROUND}

\indent To quantify hyperuniformity in scalar fields, we follow the formalism laid out in Ref. 28. For a statistically homogeneous random scalar field $F(\mathbf x)$ in $\mathbb{R}^d$ that is real valued, the autocovariance function is defined as\\
\begin{equation} \label{auto}
{\psi}(\mathbf r)=\left\langle(F(\mathbf x_1)-\left\langle F(\mathbf x_1)\right\rangle)(F(\mathbf x_2)-\left\langle F(\mathbf x_2)\right\rangle)\right\rangle,
\end{equation}
where $\mathbf r=\mathbf x_2-\mathbf x_1$, the spectral density function $\tilde\psi(\mathbf k)$ is simply the Fourier transform of the autocovariance function ${\psi}(\mathbf r)$. The hyperuniform condition is simply given by
\begin{equation} \label{hyperfield}
\lim_{|\mathbf k|\rightarrow 0} \tilde\psi(\mathbf k)=0,
\end{equation} 
which implies the following sum rule
\begin{equation} \label{sumfield}
\int_{\mathbb{R}^d} \psi(\mathbf r)d \mathbf r=0. 
\end{equation}
This means that $\psi({\bf r})$ corresponding to a homogeneous hyperuniform scalar field must be 
characterized by both positive and negative correlations such that its volume integral vanishes identically. The integrated field within a spherical window of radius $R$ will fluctuate as the window position varies. The associated variance $\sigma_F^2(R)$ is related to the autocovariance function via the relation \cite{lu1990local}\\
\begin{equation} \label{fieldvar}
\begin{aligned} 
\sigma_F^2(R)&=\frac {1} {v_1(R)} \int_{\mathbb R^d} \psi(\mathbf r)\alpha(r;R)d\mathbf r, 
%\\&=\frac {1} {v_1(R)(2\pi)^d} \int_{\mathcal R^d} \tilde\psi(\mathbf k)\tilde\alpha(k;R)d\mathbf k
\end{aligned}
\end{equation}
where $v_1(R)$ is the volume of a $d$-dimensional sphere of radius $R$, and $\alpha(r;R)$ is the scaled intersection volume, the ratio of the intersection volume of two spherical windows of radius $R$ whose centers are separated by a distance $r$ to the volume of a spherical window. For a hyperuniform scalar field, the local field variance $\sigma_F^2(R)$ decreases more rapidly than $R^{-d}$ for large $R$. \cite{PhysRevE.94.022122}\\
\indent In some instances (e.g., digitized media), it is convenient to employ a hypercubic window of side length $a$. \cite{1742-5468-2017-1-013402} Then the analog of Eq. (\ref{fieldvar}) formally applies where $R$ is replaced by $a$. Similarly, a hyperuniform scalar field is one in which $\sigma_{F}^2(a)$ decreases more rapidly than $a^{-d}$ for large $a$. Applications presented in this paper focus on discretized two-dimensional systems and so we use square windows in the subsequent sections.\\

\subsection{Level cuts of scalar fields}

\indent A two-phase random medium is a domain of space $\mathcal V \subseteq \mathbb{R}^d$ that is partitioned into two disjoint regions that make up $\mathcal V$: a phase 1 region $\mathcal V_1$ of volume fraction $\phi_1$ and a phase 2 region $\mathcal V_2$ of volume fraction $\phi_2$. The phase indicator function $\mathcal I^{(i)}(\mathbf x)$ for a given realization is defined as
\begin{equation} \label{indicator}
\mathcal I^{(i)}(\mathbf x)=\left \{ \begin{aligned} & 1, & \mathbf x \in \mathcal V_i,\\ & 0, & \mathbf x \notin \mathcal V_i, \end{aligned}\right. 
\end{equation}
Two-phase media are essential to applications as heterogeneous materials, such as composite and porous media, biological media (e.g., plant and animal tissue), foams, polymer blends, suspensions and colloids. \cite{torquato2013random} We can set a threshold $F_0$ to convert a scalar field to a two-phase medium, regions that satisfy $F(\mathbf x)>F_0$ is phase 1, and regions that satisfy $F(\mathbf x)<F_0$ is phase 2. The phase indicator function $\mathcal I(\mathbf x)$ for phase 1 is given by
\begin{equation} \label{indicator2}
\mathcal I(\mathbf x)=\Theta[F(\mathbf x)-F_0], 
\end{equation}
the phase volume fraction for phase 1 is
\begin{equation} \label{volfra}
\phi=\left\langle \mathcal I(\mathbf x) \right\rangle,
\end{equation} 
the two-point correlation function is defined as
\begin{equation} \label{s2}
S_2(\mathbf x_1,\mathbf x_2)=\left\langle \mathcal I(\mathbf x_1)\mathcal I(\mathbf x_2)\right\rangle.
\end{equation} 
This formalism can be easily generalized to the $n$-point correlation function $S_n$, \cite{torquato2013random} which is defined as
\begin{equation} \label{sn}
S_n(\mathbf x_1,\mathbf x_2,...,\mathbf x_n)=\left\langle \prod_{i=1}^n\mathcal I(\mathbf x_i)\right \rangle.
\end{equation} 
For homogeneous media, this quantity only depends on the relative displacement vector $\mathbf r_i\equiv \mathbf x_i-\mathbf x_1$. The two-point correlation function simplifies as $S_2(\mathbf x_1,\mathbf x_2)=S_2(\mathbf r)$, where $\mathbf r\equiv \mathbf x_2-\mathbf x_1$. Then the autocovariance function $\chi_{_V}(\mathbf r)$ is given by
\begin{equation} \label{chiv}
\chi_{_V}(\mathbf r)\equiv S_2(\mathbf r)-\phi^2.
\end{equation}
\indent The hyperuniformity condition for two-phase media is given by the following spectral-density
condition:
\begin{equation} \label{hyper2p}
\lim_{|\mathbf k|\rightarrow 0} \tilde\chi_{_V}(\mathbf k)=0.
\end{equation}
This implies the following sum rule in direct space \cite{0953-8984-28-41-414012}
\begin{equation} \label{sum2p}
\int_{\mathbb{R}^d} \chi_{_V}(\mathbf r)d \mathbf r=0. 
\end{equation}
The hyperuniformity of two-phase media obtained from level cuts of scalar fields is discussed in Sec. V. 

\section{GENERATION OF SCALAR FIELDS}

\subsection{Gaussian Random Fields}

Gaussian random fields have been used to model a variety of systems, including microwave background radiation, \cite{liddle2000cosmological} heterogeneous materials, \cite{roberts1995transport} and random speckle fields. \cite{dainty1977statistics} A Gaussian random field $f(\mathbf r)$ is constructed by a superposition of plane waves with random wave vectors and phases, \cite{berk1991scattering} i.e.,
\begin{equation} \label{Gaussian}
f(\mathbf r)=\frac {1} {\sqrt{N}} \sum\limits_{i=1}^{N} {A_i} \cos(\mathbf k_i \cdot \mathbf r+\phi_i),
\end{equation}
where direction of $\mathbf k_i$ is uniformly randomly oriented and the phase $\phi_i$ is a random number that lies uniformly in the interval $[0,2\pi]$. In the large-$N$ limit, the central limit theorem ensures a Gaussian distribution of the field. Since it is widely used to model multiphase heterogeneous materials, it provides a good starting point to look for hyperuniformity.\\ 
\indent Without loss of generality, we normalize the field so that the autocovariance at the origin is unity, i.e., $\psi({\mathbf r=\mathbf 0})=1$. Then $A_i$ satisfies
 \begin{equation} \label{discrete}
 \frac {1}{2N}\sum A_i^2=1. 
 \end{equation}
In the continuous limit, we have \cite{dennis2007nodal}
\begin{equation} \label{continous}
\psi(\mathbf r=\mathbf 0)=\int_{\mathbb{R}^d} \tilde\psi(\mathbf k) d\mathbf k=1, 
\end{equation} 
where $\tilde\psi(\mathbf k)$ is the spectral density. For Gaussian variables with zero mean, the two-point information encoded in $\psi(\mathbf r)$ totally determines any higher-order correlation function via the correlation matrix. \cite{dennis2007nodal}\\ 

\subsection{Spinodal Decomposition}

\indent The Cahn-Hilliard equation was introduced to describe phase separation by spinodal decomposition almost sixty years ago. \cite{cahn1958free} It has been applied to model alloys, \cite{rundman1967early} polymer blends \cite{smolders1971liquid} and even pattern formations in ecology. \cite{liu2013phase} The equation was derived from the expansion of free energy regarding to local concentration $c(\mathbf x,t)$ in an inhomogeneous liquid that incorporates the energy of a homogeneous one and a gradient term which penalizes the existence of concentration gradients. The free energy can be written as a functional of the form
\begin{equation} \label{free}
\mathcal{F}[c]=\int [f(c)+\frac {\gamma}{2}{|\nabla c|}^2]d\mathbf x,
\end{equation}
where $c(\mathbf x,t)$ is the concentration of the fluid at position $\mathbf x$ and time $t$, $f(c)$ is the free energy density of a homogeneous material. Normally, the latter quantity is written in a double-well form, e.g., in this paper, we use the form $f(c)=\frac {1} {4} (c^2-1)^2$. It can be seen that $c=\pm1$ corresponds to two different phases. Finally, the parameter ${\gamma}$ controls the width of the transition regions between different phases.\\ 
\indent The time evolution of the concentration field $c(\mathbf x,t)$ is described by the equation
\begin{equation}  \label{ch}
\frac{\partial{c}}{\partial{t}}=\nabla \cdot(D\nabla \mu),
\end{equation}
where $D$ is the diffusion coefficient, and $\mu=f^{\prime}(c)-\gamma {\nabla}^2c$ is the chemical potential. One can see that $\sqrt{\gamma}$ sets the width of the transition regions between different phases by solving the equation in 1D in equilibrium, the solution is simply $c(x)=\tanh({x\over \sqrt{2\gamma}})$. \\
\indent Typically, one begins with random initial conditions, which means that the binary liquid is well-mixed and follows the evolution of $c(\mathbf x,t)$ as governed by Eq. (\ref{ch}). The more evenly mixed, the more accurate the following arguments hold. In experiments, this can be achieved by a rapid quench, splat cooling or vapor quenching. \cite{cook1970brownian} Then domains of two phases will emerge and go through a coarsening process. During the early stage, it was suggested that Brownian diffusion is important. \cite{cook1970brownian} However, at later times, the system will enter a so-called ``scaling regime'', which means the system will show self-similarity. Although the domain size will grow continuously following a power law $t^{\frac {1} {3}}$, the system looks statistically the same. \\ 
\indent In the scaling regime, the autocovariance function and spectral density function will collapse to the same curves respectively after being rescaled to the scaling functions $\mathscr{G}(x)$ and $\mathscr{F}(x)$. Specifically, in 2D we have\\
\begin{equation} \label{psiscaling}
{\psi}(r,t)=\mathscr{G}(rk_1(t)),
\end{equation}
and,
\begin{equation} \label{specscaling}
\mathscr{F}(k/k_1(t))=k_1(t)^2\tilde\psi(k,t),
\end{equation}
where $\psi(r,t)$ is defined as Eq. (\ref{auto}) with respect to the concentration field $c(\mathbf x,t)$, $k_1(t)$ characterizes the typical length scale and there are several different ways to define it. Normally, the spectral density function $\tilde\psi(k,t)$ here is actually normalized by $\left\langle c^2\right\rangle-{\left\langle c\right\rangle}^2$, since multiplying a constant would not change anything regards to the determination of hyperuniformity, in order to avoid confusion, we still used the same notation here. There are several ways to determine $k_1(t)$, such as the peak location in the spectral density function, the first moment of the spectral density function, and the first zero in the autocovariance function. Here the first-moment method is used, \cite{zhu1999coarsening} i.e., 
\begin{equation} \label{firstmoment}
k_1(t)=\frac {\Sigma k\tilde\psi(k,t)} {\Sigma \tilde\psi(k,t)}.
\end{equation}       
\indent It is noteworthy that Furukawa \cite{furukawa1984dynamics} has claimed that at deep quenches, where the thermal force contributes less significantly, the spectral density can have a form such that it goes to zero as a power law ($\sim k^2$); see also. \cite{bray1993theory} (Note that all of our simulations are carried out only for 2D; real experiment would be carried in 3D and so the scalings should be different). Some experiments seem to support this scaling behavior, \cite{hashimoto1986late} but the fitting for small $k$ was actually done in a region immediately before the peak; see Fig. 1. The material used is the mixture of PM and PVME and it took one to two hours to get to the scaling regime. It is clear that the smallest $k$ they obtained is as large as $0.5k_{peak}$, which is not small enough to make any statement about the behavior when $k\rightarrow 0$.\\    
\begin{figure}[H] 
\centering
\includegraphics[width=8cm,height=6cm]{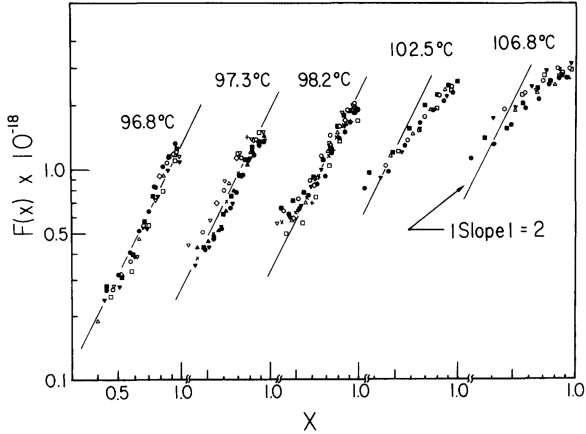}
\caption{The asymptotic behavior of the experimental scaled structure factors (symbols) and the theoretical scaled structure factors (solid lines) at small $k$, where $F(x)$ is the scaled structure factor and $x$ is $k/k_{peak}$. We can see that at small $k$, $F(x) \sim x^2$. Reproduced from, \cite{hashimoto1986late} with the permission of AIP Publishing.}
\end{figure}

\subsection{Swift-Hohenberg equation}

\indent Swift-Hohenberg equations were developed to study Rayleigh-B{\' e}nard (RB) convection in hydrodynamics. Such equations involve a stochastic thermal noise term to account for the effect of fluctuations. However, the deterministic version is of great importance in nonlinear pattern formation by itself, and, often, fluctuations can be neglected. For our purposes, here we only investigate the deterministic version of the Swift-Hohenberg equation, which takes the form \cite{cross1993pattern}
\begin{equation} \label{sh}
\frac{\partial{u}}{\partial{t}}=\epsilon u-(\nabla^2+1)^2 u-u^3,
\end{equation}  
where $u(\mathbf x,t)$ is in general a real field in d-dimensional space and the autocovariance function $\psi(r,t)$ is defined as Eq. (\ref{auto}) with respect to the field $u(\mathbf x,t)$. \\
\indent The Swift-Hohenberg equation and its variants can give rise to crystal \cite{cross1993pattern} and quasicrystal patterns. \cite{PhysRevLett.79.1261} However, the most common patterns that arise are labyrinth-like textures that consist of ``channels" whose widths are determined by a pre-selected wavelength. To make this feature apparent, the equation that we actually will solve is a transformation of Eq. (\ref{sh}), the form of which emphasizes the selected wavenumber $k_0$, \cite{hu2005statistical} i.e.,
\begin{equation} \label{sh2}
\frac{\partial{u}}{\partial{t}}=D[\epsilon-(\nabla^2+{k_0}^2)^2]u-u^3,
\end{equation}
where $D$ is another parameter that sets the rate of diffusion. The transformation leading to Eq. (\ref{sh2}) is $x=k_0x_k$, $\epsilon=\epsilon_k/k_0^4$, $t=Dk_0^4t_k$ and $u=u_k/D^{1/2}k_0^2$, here the subscript $k$ denotes the case of Eq. (\ref{sh2}).\\
\indent The hyperuniformity of a certain labyrinth-like Turing patterns has been analyzed by Torquato, \cite{0953-8984-28-41-414012} where the pattern was thresholded to make a two-phase system with a certain characteristic wavelength. (Turing proposed a reaction-diffusion theory of morphogenesis \cite{turing1952chemical} that has served to model stripes, spots and spirals that arise in biology). The spectral density of the thresholded image shows a well-defined annulus in which the scattering intensity is enhanced, but the system is not hyperuniform. This counterintuitive result is consistent with the fact that dense packings of spheres of identical size are not necessarily
hyperuniform. \cite{zhang2013precise} These results show that hyperuniformity is very subtle because it demonstrates that assembling space-filling "building blocks" of fixed size does not guarantee the anomalous suppression of large-scale density fluctuations associated with hyperuniformity. The aforementioned finding about Turing patterns also leads to the natural theoretical question of whether there exists hyperuniform two-phase systems with spectral densities in which scattering is entirely concentrated within some relatively thin annulus defined by a small range of wavenumbers away from the origin. It was proved \cite{0953-8984-28-41-414012} that a ring-like spectral density is not realizable for two-phase media for it implies a vanishing specific surface. However, if we consider the system in ``grey-scale", which means treating it as a scalar field, this analysis does not apply. In the next section, we will demonstrate that these labyrinth-like patterns are effectively hyperuniform, or even ``stealthy" \cite{batten2008classical, torquato2015ensemble, zhang2015ground} in the same sense of Eq. (\ref{stealthy}), but with $S(\mathbf k)$ replaced by $\tilde\psi(\mathbf k)$.\\      

\section{SIMULATION PROCEDURES AND RESULTS}

\subsection{Hyperuniformity and Gaussian random fields}

\indent From the equations presented in Sec. III, it is clear that Gaussian random fields are constructed by summing up elements with wave vectors ${\bf k}_i$ such that the sum 
rule Eq. (\ref{continous}) on the autocovariance function in the continuous limit is satisfied. Thus, constructing hyperuniform scalar fields should be straightforward. In the continuous limit, Eq. (\ref{Gaussian}) becomes an integral in $\bf k$-space, we then can choose $A_i$ as a function of $k$ that vanishes at the origin. (This is due to the fact that the spectral density is proportional to $A^2(k)$, and the only constraint is that the volume integral of $A^2(k)$ converges.) Numerically, one can only do the integral in digitized $\bf k$-space for a finite size system, however, one can always improve the resolution to meet any practical requirement. The general formula can be written as\\
\begin{equation} \label{construct}
f(\mathbf r)=C \sum\limits_{i=1}^{N} {A({k_i})} \cos(\mathbf k_i \cdot \mathbf r+\phi_i) \Delta k_i^d
\end{equation}
where $C$ is just a normalization constant, and $\Delta k_i^d$ is the volume element in $d$ dimensions that contains $\mathbf k_i$.  
Here two choices of $A(k)$ in 2D are presented:\\   
\begin{equation} \label{exp}
A({k})=\sqrt{k}e^{-\alpha k^2}.
\end{equation}
\begin{equation} \label{power}
A({k})=\frac {k^2}{\sqrt{\beta+k^{10}}}.
\end{equation}

\begin{figure}[H]
\centering
\subfigure[]{
\includegraphics[width=6.4cm, height=6cm]{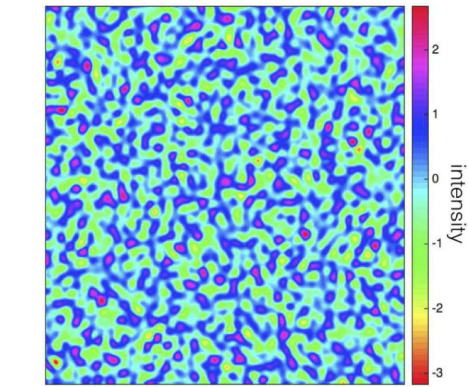}
}
\subfigure[]{
\includegraphics[width=6.4cm, height=6cm]{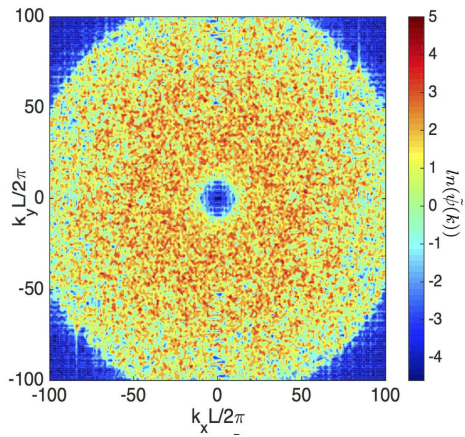}
}
\caption{(Color online.) (a) Realization of a stealthy scalar field corresponding to Eq. (\ref{exp}) with $\alpha=2.5$ and $K=10\Delta k$. There are 10,000 plane waves added together with sample size $1000\times1000$. Here only a $300\times300$ portion is shown for clarity. (b) The corresponding spectral density of (a), the color bar is in logarithm scale. The spectral density has a ``hole" in the origin.}
\end{figure}
\noindent where $\alpha$ and $\beta$ are two positive constants. For both cases, $A(k)$ vanishes at the origin and the volume integral of $A^2(k)$ converges. The first choice Eq. (\ref{exp}) has an exponential decay tail and the corresponding spectral density goes to zero linearly at the origin. The second choice Eq. (\ref{power}) has a power-law decay and the corresponding spectral density goes to zero quartically at the origin. Each one is a superposition of 10,000 plane waves ($N=10000$), which are grouped evenly into 100 groups from $\Delta k$ to $100\Delta k$ with random orientations, where $\Delta k$ is simply $2\pi/L$, where $L$ is the side length in the simulation. Since polar coordinates are used here, $\Delta k_i^2$ will contribute factors $k_i$ and $\Delta k$. The exact value of $C$ is irrelevant to hyperuniformity so the summation is not normalized.\\

\begin{figure}[H]
\centering
\subfigure[]{
\includegraphics[width=6.4cm, height=6cm]{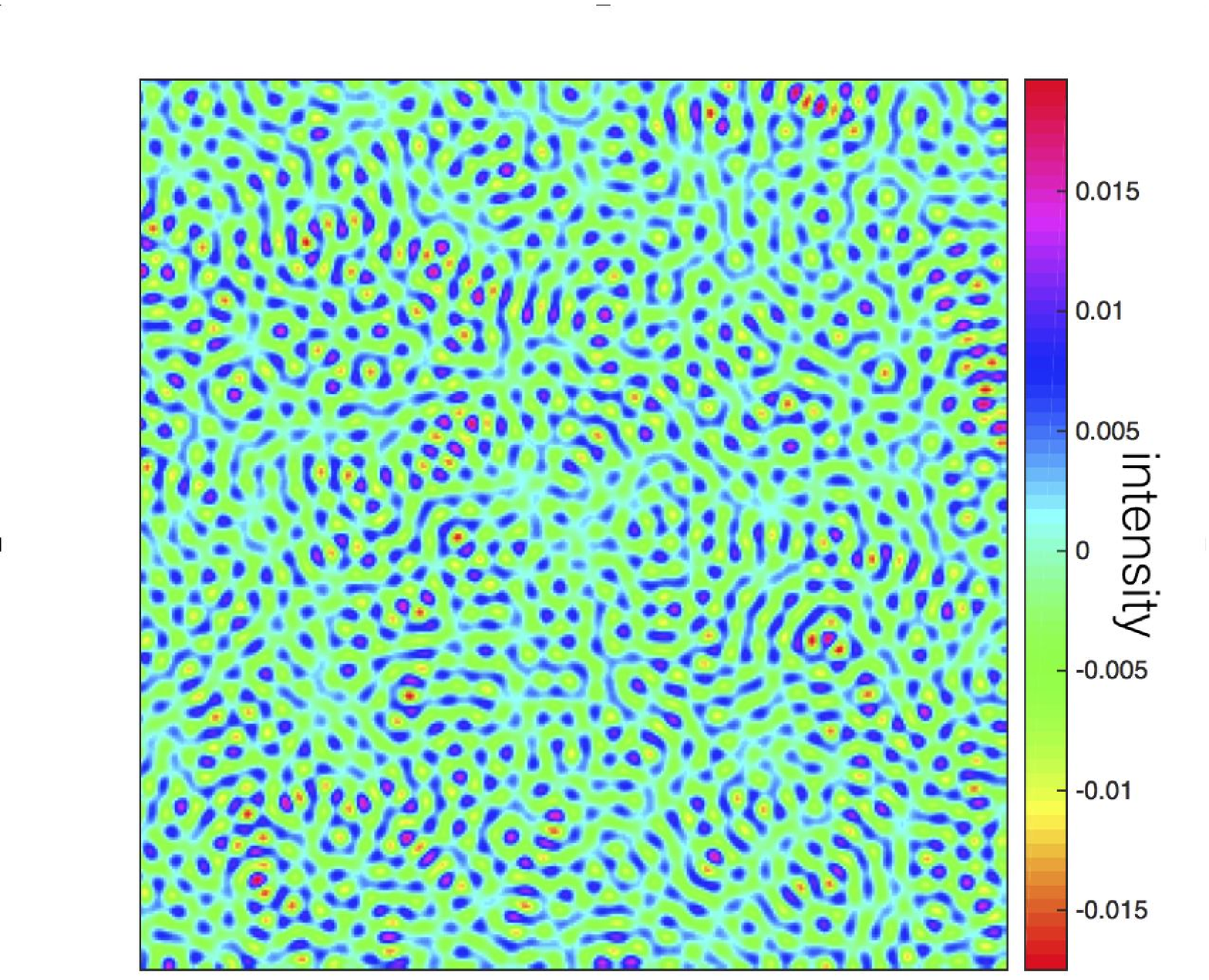}
}
\subfigure[]{
\includegraphics[width=6.4cm, height=6cm]{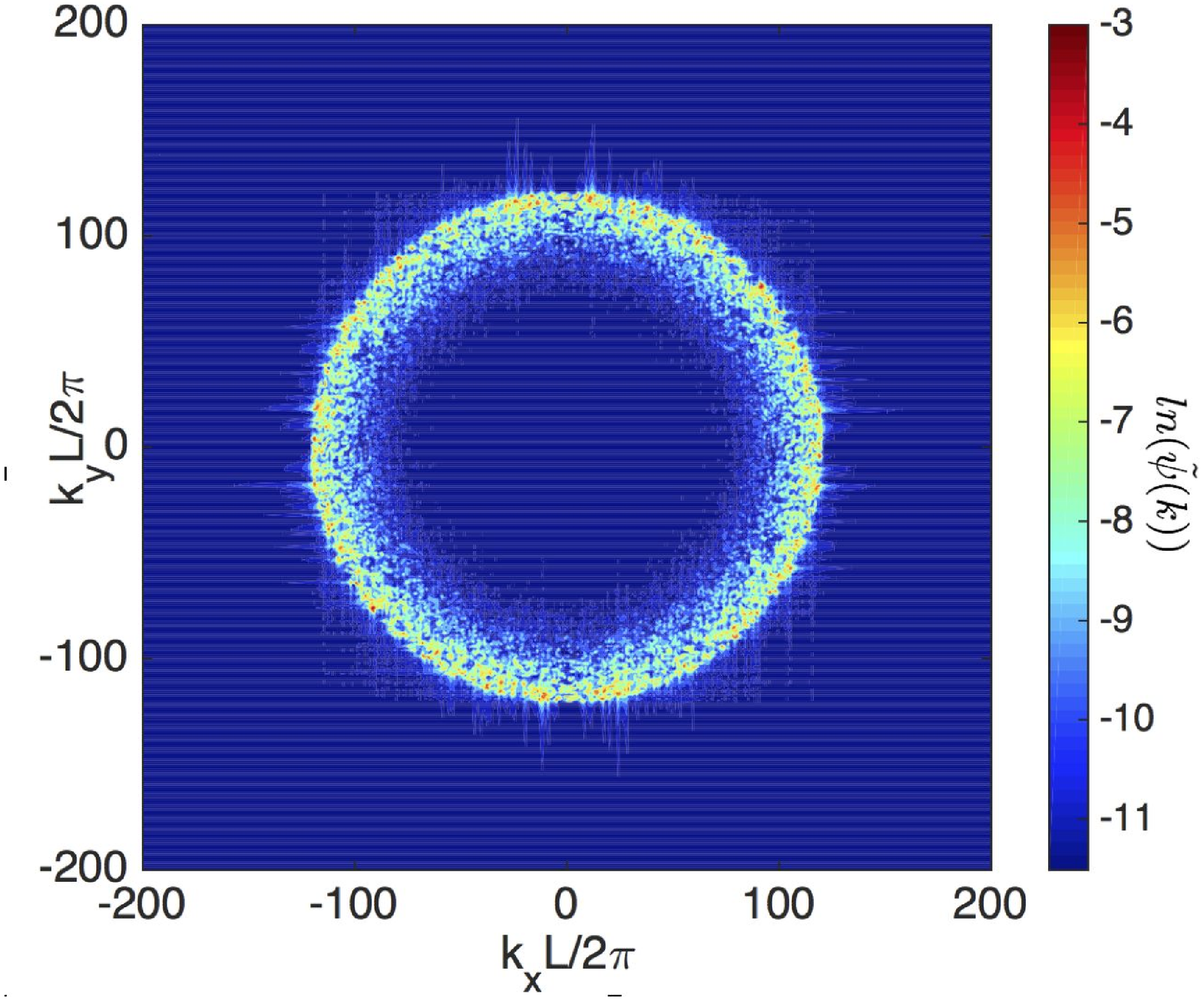}
}
\caption{(Color online.) (a) Realization of a stealthy scalar field corresponding to Eq. (\ref{power}) with $\beta=0.00001$ and $K=10\Delta k$. There are 10,000 plane waves added together with sample size $1000\times1000$. Here only a $300\times300$ portion is shown for clarity. (b) The corresponding spectral density of (a), the color bar is in logarithm scale. The spectral density shows a sharp ring.}
\end{figure}

\indent It is very straightforward to transform these systems to stealthy ones by simply applying a translation operation to the spectral density so that $\tilde\psi_{stealthy}(k)=\tilde\psi_{hyperuniform}(k-K)$, for $k\geq K$ and $\tilde\psi_{stealthy}(k)=0$, for $k<K$. This operation applied to Eq. (\ref{construct}) yields\\
\begin{equation} \label{construct2}
f(\mathbf r)=C \sum\limits_{i=1}^{N} {A({k_i})}  \cos((\mathbf k_i +K\hat{\mathbf k_i})\cdot \mathbf r+\phi_i) \Delta k_i^d.
\end{equation}

\indent Realizations of stealthy scalar fields corresponding to Eq. (\ref{exp}) and Eq. (\ref{power}) ($K$ is chosen to be $10\Delta k$) as well as the spectral densities themselves are shown in Figs. 2 and 3 for certain parameters. These two constructions exhibit different morphologies, but each are characterized by an exclusion region around the origin $\bf k=0$ in which the intensities are
very small. Indeed, the hyperuniformity metric \cite{atkinson2016critical}
\begin{equation} \label{metric}
H \equiv \frac {\tilde \psi(0^+)} {{\tilde \psi(k_{peak})}}
\end{equation}
is relatively very small ($\sim 10^{-4}$), where $k_{peak}$ is the wavenumber at the peak height of the spectral function. The hyperuniformity metric $H$ provides a measure of the degree to which density
fluctuations are suppressed at large length scales. A perfectly hyperuniform system has $H=0$, but a system may be considered to be effectively hyperuniform when $H$ is of the order of $10^{-4}$ or smaller. \cite{atkinson2016critical}\\
\indent Our results show that Gaussian random fields provide a very straightforward way to generate targeted disordered hyperuniform or stealthy scalar fields at very large length scales and thus can guide experimentalists to synthesize new hyperuniform materials via stereolithography
and/or 3D printing techniques. The question of whether two-phase media obtained by thresholding hyperuniform Gaussian random fields can be hyperuniform will be answered in Sec. V.\\     

\subsection{Hyperuniformity and the Cahn-Hilliard equation}

\indent Here we simulate the evolution of the concentration field using the Cahn-Hilliard equation (\ref{ch}). The parameters used in our simulation are as following: $D=0.01, {\gamma}=0.5$. The initial condition is a random scalar field with the amplitude 0.0001 (can be both positive and negative). The equation was integrated on a 1000$\times$1000 2D grid under periodic boundary conditions. The grid spacing is unity, and time step is unity (with a characteristic time $\tau=\gamma/{2D}=25$). Patterns at different times are shown in Fig. 4 below.\\
\begin{figure}[H]
\centering
\subfigure[\ t=10000]{
\includegraphics[width=3.5cm, height=3.5cm]{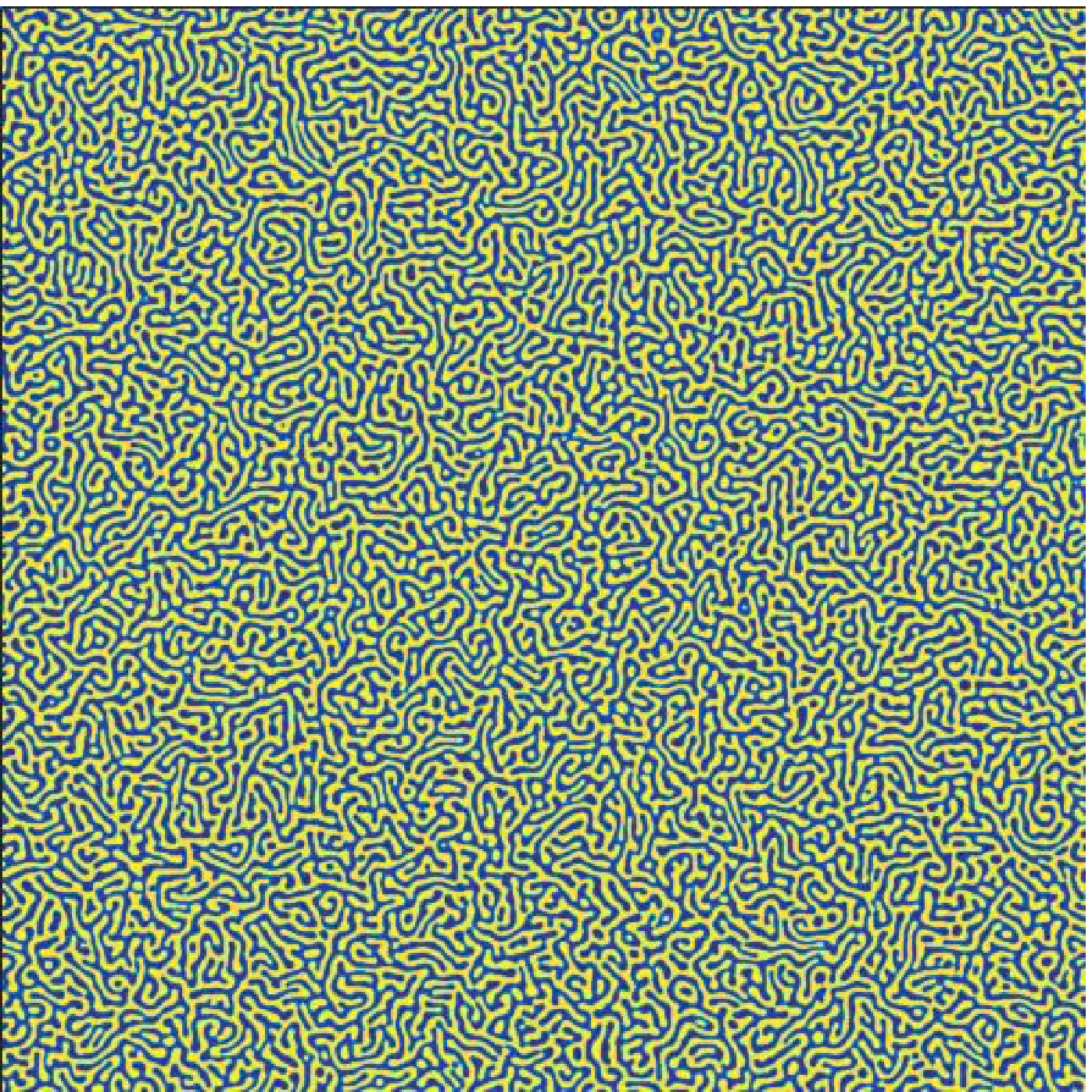}
}
\subfigure[\ t=100000]{
\includegraphics[width=3.5cm, height=3.5cm]{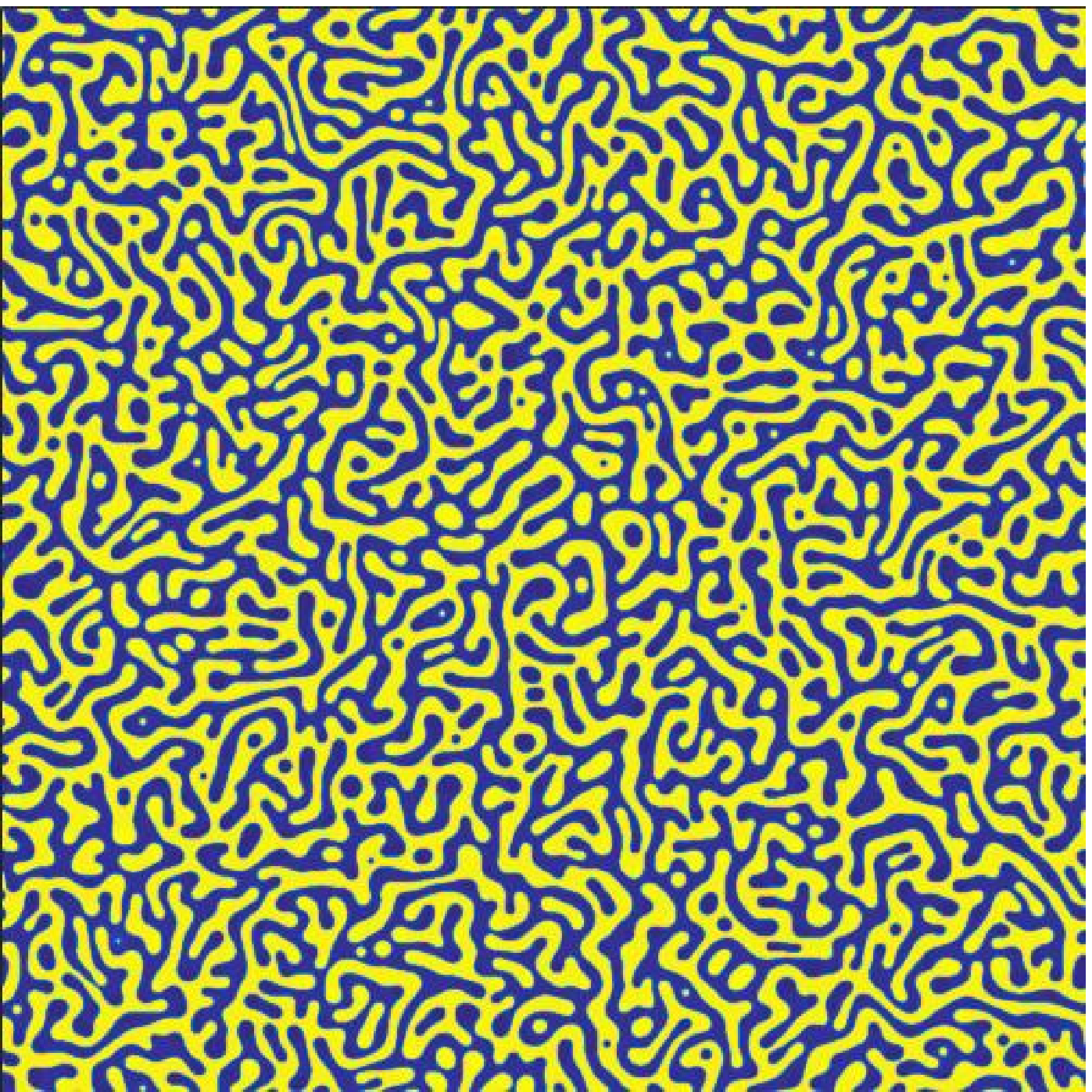}
}
\subfigure[\ t=200000]{
\includegraphics[width=3.5cm, height=3.5cm]{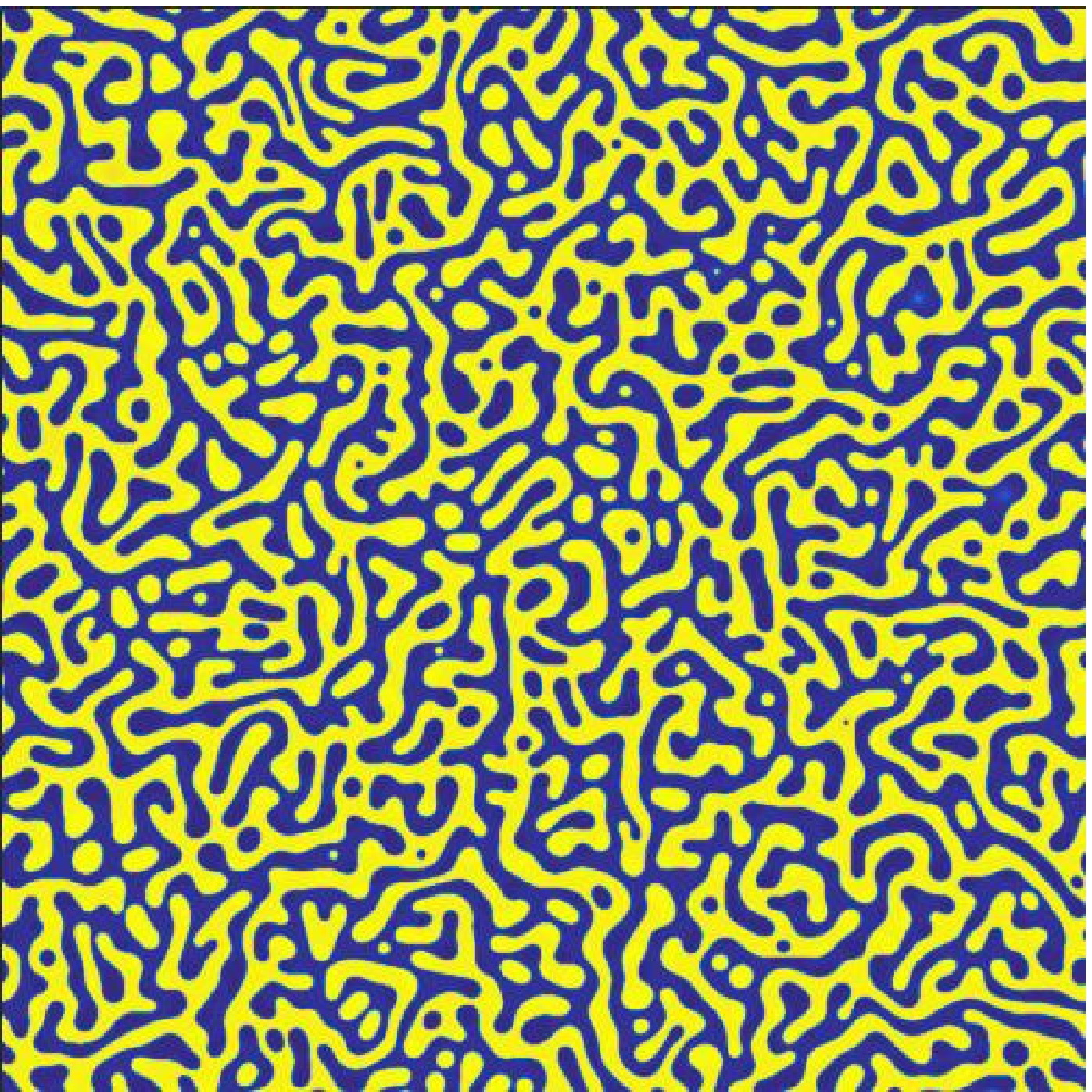}
}
\subfigure[\ t=1000000]{
\includegraphics[width=3.5cm, height=3.5cm]{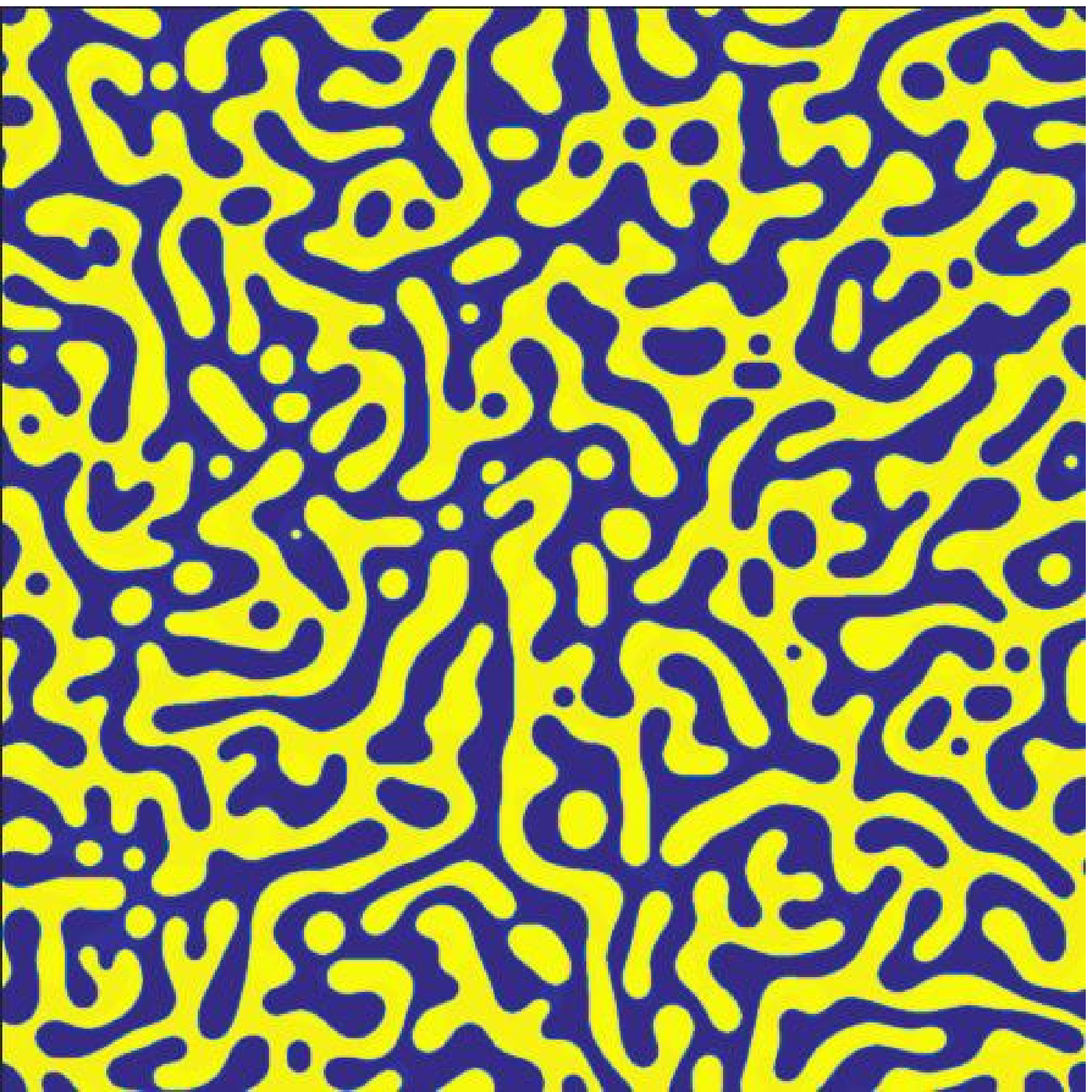}
}
\caption{(Color online.) The time evolution of the concentration field of a system evolving under the Cahn-Hilliard equation (\ref{ch}) with critical quench (volume fraction ratio of the two phases is 1:1), four snapshot taken at t=10,000, t=100,000, t=200,000 and t=1,000,000 are shown here, the system size is 1000$\times$1000, the characteristic time $\tau=\gamma/{2D}=25$.}
\end{figure}

\begin{figure}[H]
\centering
\subfigure[\ t=10000]{
\includegraphics[width=3.5cm, height=3.5cm]{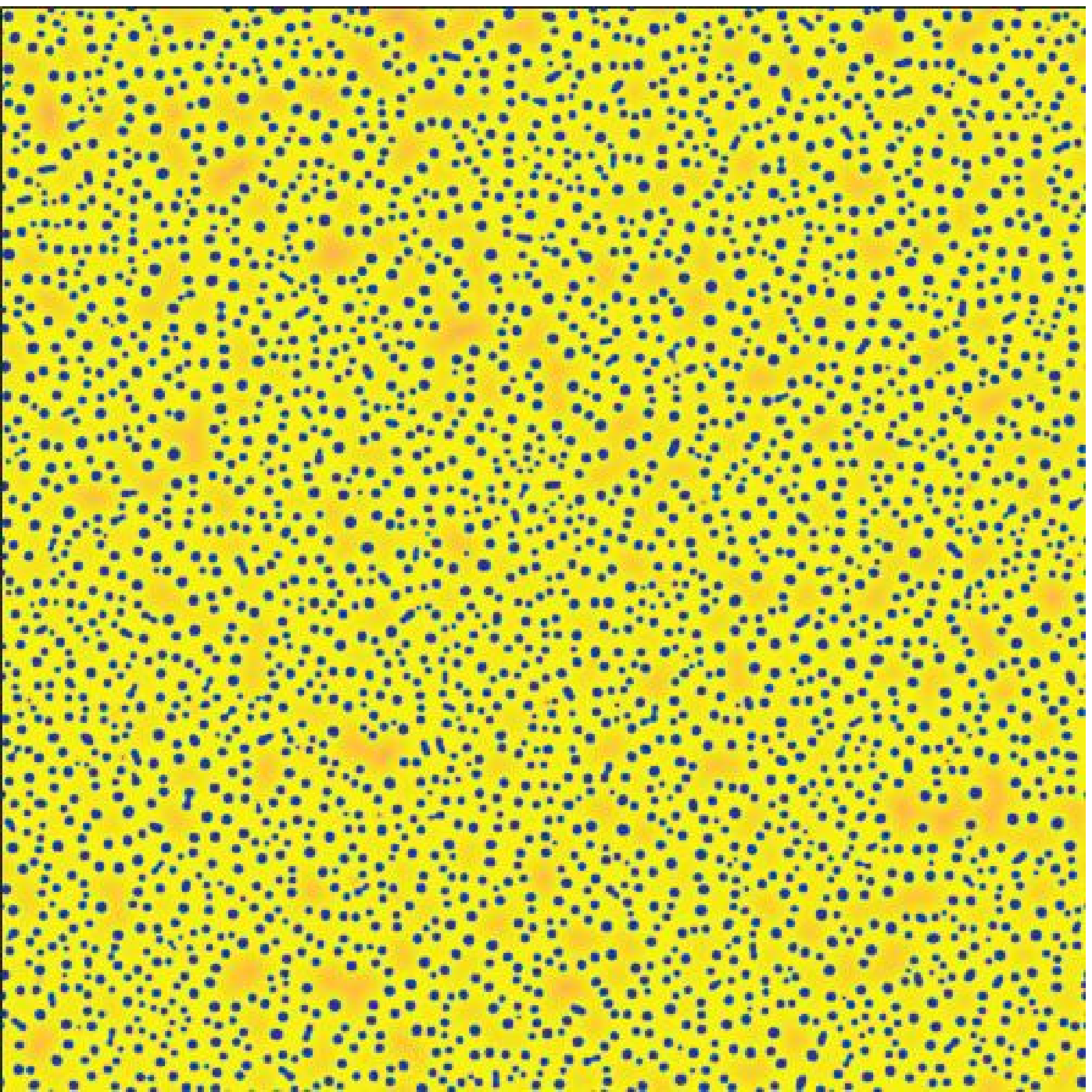}
}
\subfigure[\ t=100000]{
\includegraphics[width=3.5cm, height=3.5cm]{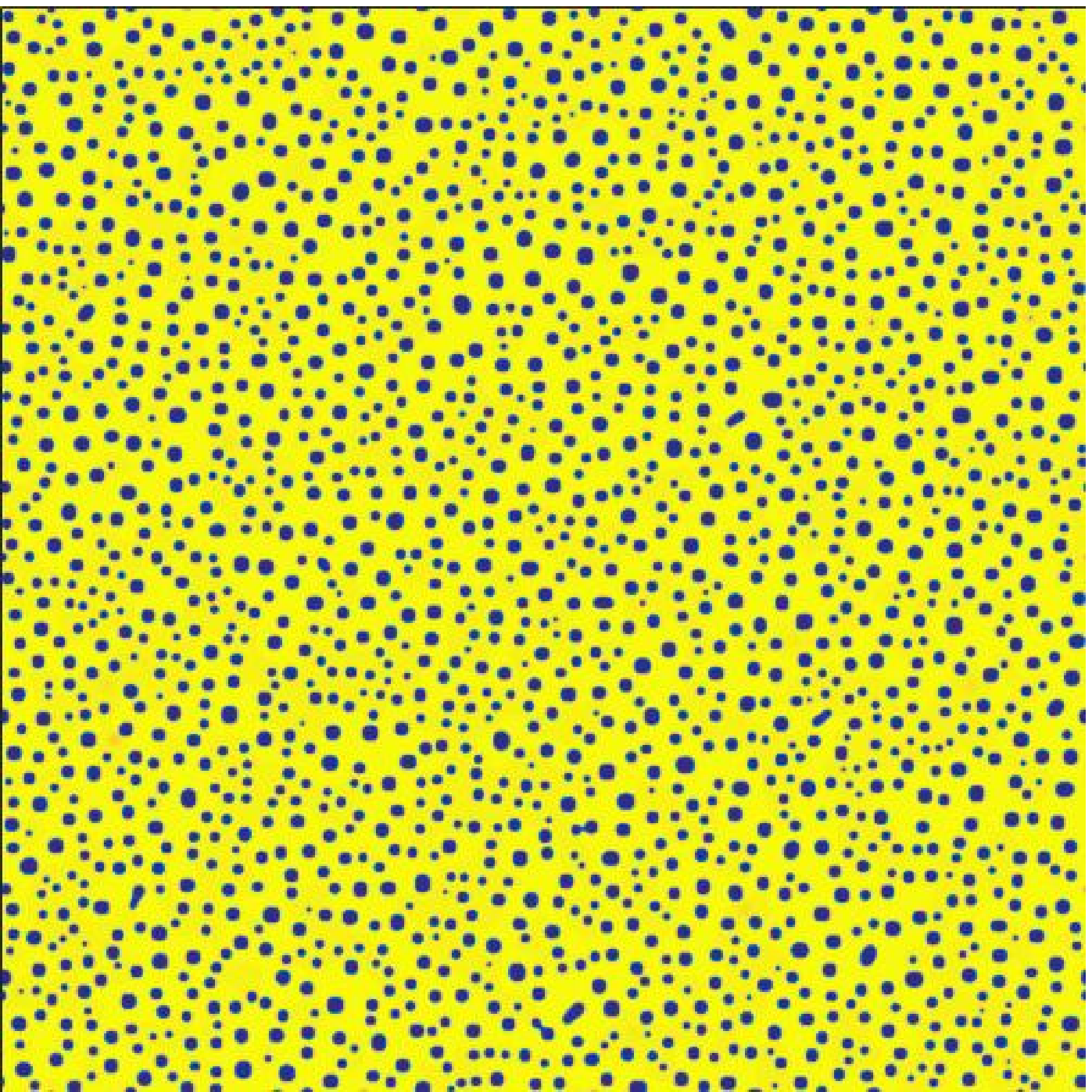}
}
\caption{(Color online.) The time evolution of the concentration field of a system evolving under the Cahn-Hilliard equation (\ref{ch}) off critical quench (volume fraction ratio of the blue phase and yellow phase is 2:8), two snapshot taken at t=10,000 and t=100,000 are shown here, the system size is 1000$\times$1000, the characteristic time $\tau=\gamma/{2D}=25$.}
\end{figure}

\indent Under this set up, the system is in a critical quench, which means that the two phases have equal volumes. This can be changed by using a biased initial condition, if one phase dominates in volume fraction, then it will form droplets, which is shown in Fig. 5. Here we focus on the analyses of the critical quench case. An example of the autocovariance function is shown in Fig. 6, which shows that it decays very fast near the origin but apparently has a long tail. The evolution of the spectral density is shown in Fig. 7. It is seen that after scaling, all of the curves nearly collapse on to a single scaling curve after t=100,000.\\

\begin{figure}[H]
\centering
\includegraphics[clip,width=8cm,height=6cm]{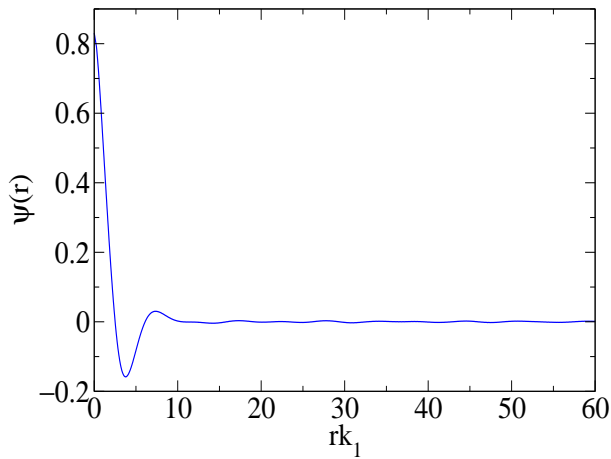}
\caption{(Color online.) The autocovariance function $\psi(r)$ versus $rk_1$ at t=100,000 associated with the right-upper panel of Fig. 4, where $r$ is scaled by the characteristic wavenumber $k_1$. The characteristic time $\tau=\gamma/{2D}=25$.}
\end{figure}
\begin{figure}[H]
\centering
\includegraphics[width=8cm,height=6cm]{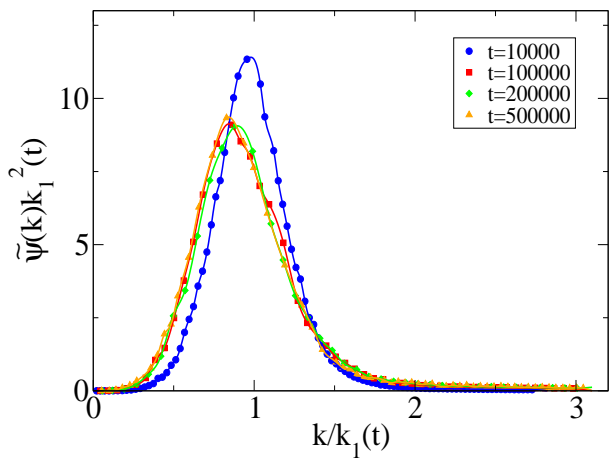}
\caption{(Color online.) The scaled spectral density $\tilde \psi(k)k_1^2(t)$ versus $k/k_1(t)$ at t=10,000, t=100,000, t=200,000 and t=500,000 associated with Fig. 4 (the pattern at t=500,000 is not shown), all curves collapse on to one another after t=100,000. The characteristic time $\tau=\gamma/{2D}=25$.}
\end{figure}
\indent Hyperuniformity in these systems is revealed by scalar field fluctuations (denoted by $\sigma_F^2(a)$) as well as the behavior of the spectral density near the origin. 
It is seen that the scalar field fluctuations decay as $a^{-3}$, which is faster than $a^{-2}$, the case of typical random scalar fields in 2D; while the log-log spectral density near the origin shows it goes to zero as $k^{4}$, which is consistent with the exponent found in scalar field fluctuations. In order to minimize finite size effects, the results here are obtained for a very large system (10000$\times$10000) at time t=100,000; See Figs. 8 and 9.\\
\begin{figure}[H]
\centering
\includegraphics[width=8cm,height=6cm,clip=]{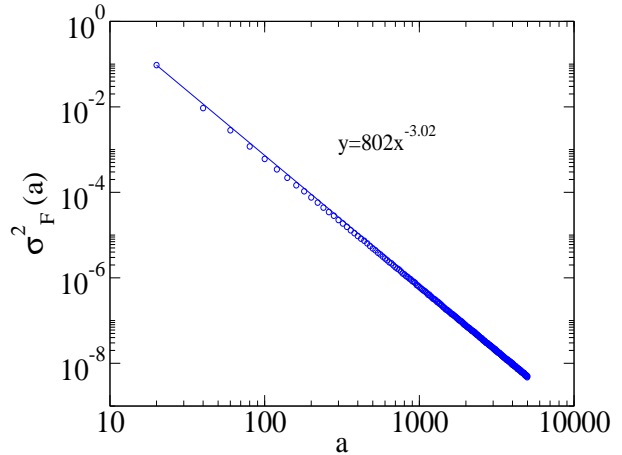}
\caption{(Color online.) Scalar-field fluctuations $\sigma_F^2(a)$ as a function of the window side length $a$ at t=100,000 associated with a 10000$\times$10000 system. The largest window shown here is 5000$\times$5000.}
\end{figure}
\begin{figure}[H]
\centering
\includegraphics[width=8cm,height=6cm,clip=]{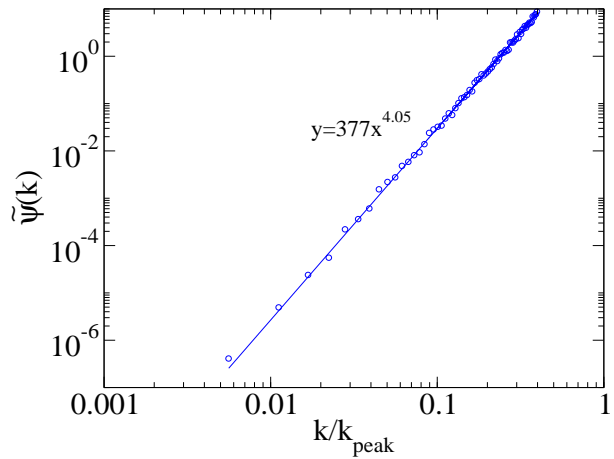}
\caption{(Color online.) Spectral density $\tilde \psi(k)$ as a function of the ratio $k/k_{peak}$ at t=100,000 associated with a 10000$\times$10000 system, where $k_{peak}$ denotes the position of the peak. It is shown that the spectral density goes to zero quartically in $k$ ($k^4$).}
\end{figure}
\indent Since the system has already entered the scaling regime, we would get the same scaling behavior for scalar field fluctuations at different times, which also holds true for spectral densities at small $k$. Thus it is safe to say that the system will remain hyperuniform for all future times in the infinite-size limit. For finite systems, they will finally break up into two large disjointed domains and lose the bicontinuous structure, which implies a proper time window is required to observe this phenomenon experimentally. \\ 
\indent Although our calculations focused on critical quenches in the scaling regime, there are two more observations that are noteworthy as well. From Fig. 7, it is seen that when t=10,000, the system is not in the scaling regime. However, the spectral density function at t=10,000 actually has a higher peak compared to those in the following times, and a careful look at the low-$k$ region shows that low-$k$ values are smaller than the ones obtained from the scaling regime. Namely, for the same $k/k_{peak}$ ratio, $\tilde\psi(k_{peak})/\tilde\psi(k)$ is higher, and hence the hyperuniformity metric $H$ [defined by Eq. (\ref{metric})] is smaller. In this sense, the system at t=10,000 is even more hyperuniform, which implies that the system may already be hyperuniform before entering the scaling regime. This means that experimentally one may not need to wait till the system enters the scaling regime to prepare a hyperuniform state.\\
\indent We also investigated the scalar field fluctuations and the spectral density function for systems off critical quench, not very surprisingly the scalar field fluctuations have the same scaling as $a^{-3}$. However, the scaling of the spectral density is more sensitive and deviates at the first few smallest wave numbers. We believe this discrepancy is induced by the fact that the interfacial region is more important for droplet patterns and a limited resolution of the mesh can impair the hyperuniformity. We find that an increase in the resolution and the system size do result in more hyperuniform spectral densities. We expect that systems off the critical quench are still hyperuniform. What is remarkable about these systems is that the phase with smaller volume fraction consists of numerous droplets with a certain distribution of sizes, which is reminiscent of previous work done for hyperuniform point configurations \cite{torquato2015ensemble} and spherical packings, \cite{zachary2011hyperuniformity} despite the fact that here we are dealing with a scalar field. Nonetheless, an analog to particle systems may help us think why spinodal decomposition patterns are hyperuniform. From the patterns of droplets (Fig. 5), we can see that two neighboring droplets tend to maintain some distance between each other; otherwise they will merge into a single droplet. This effective short-ranged interaction between droplets is beneficial to establish large-scale order in these systems, just like it does in jammed packings. \cite{torquato2000random, torquato2010jammed} Moreover, the domain size grows slower ($t^{\frac {1}{3}}$) than the diffusion rate ($t^{\frac {1}{2}}$), which tends to give the system enough time to wipe out the inhomogeneity inside each domain. The configurations shown in Fig. 5 are essentially polydisperse packings of disks (or spheres in 3D) and should be of great importance in applications, since such configurations can be generated directly through the integral of a time evolution PDE rather than time consuming optimization algorithms. Moreover, they could also be realized easily by modern fabrication techniques such as 3D printing.\\

\subsection{Hyperuniformity and the Swift-Hohenberg equation}
\indent Here we simulate pattern formation under the Swift-Hohenberg equation Eq. (\ref{sh2}) and analyze its degree of hyperuniformity. We restrict ourselves to the parameter space that gives rise to labyrinth-like patterns.\\
\indent All results shown here are obtained with parameters $D=0.01$ and $\epsilon=0.1$, and $k_0$ is varied to achieve different characteristic wavelengths $2\pi/k_0$ and different degrees of disorder. The initial condition is a random scalar field uniformly distributed from -0.001 to 0.001. The equation was integrated on a $1000\times1000$ 2D grid under periodic boundary conditions. The grid spacing is unity, and the time step is also unity. The desired patterns are almost formed at t=10,000, and calculations shown here are all done at t=100,000.\\
\indent A pattern at t=100,000 with wave vector $k_0=0.7$ is presented in Fig. 10. The evolution of the spectral densities of this system are shown in Fig. 11. It is noticeable that a peak emerges at the selected wave vector and fully forms there at large times. The hyperuniformity metric $H$ is between $10^{-5}$ and $10^{-4}$ at t=100,000, which is considered to be effectively hyperuniform. \cite{atkinson2016critical} But the situation here is more subtle than in the case of spinodal decomposition. The spectral density drops very quickly away from the peak. It also is essentially zero for a continuous range of $k$ from $k=0$
to almost its peak value, and hence becomes stealthy-like (see Sec. III). Compared to randomly jammed monodisperse particles, \cite{torquato2000random} these disordered labyrinth-like patterns are ``jammed" in the sense that the channels are of equal width and they partition the space completely. This simple observation may explain why these patterns are hyperuniform.\\
\begin{figure}[H]
\centering
\includegraphics[width=7cm,height=6.3cm]{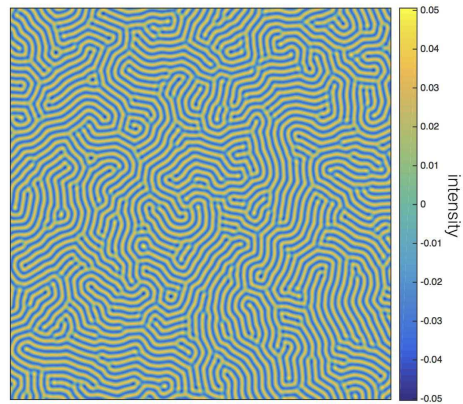}
\caption{(Color online.) A typical pattern under the dynamics of the Swift-Hohenberg equation with the selected wave vector $k_0=0.7$ at t=100,000. The system size is 1000$\times$1000. Here a 500$\times$500 portion is shown.}
\end{figure}

\begin{figure}[H]
\centering
\includegraphics[width=8cm,height=6cm]{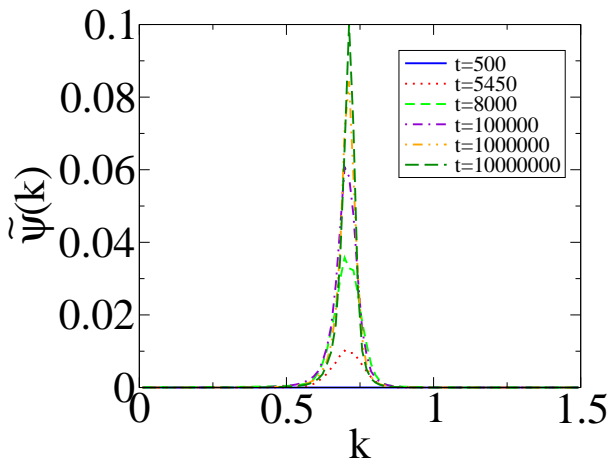}
\caption{(Color online.) The time evolution of the angular averaged spectral density $\tilde \psi(k)$ as a function of $k$ associated with a 1000$\times$1000 system with the selected wave vector $k_0=0.7$. Spectral densities are shown at t=500, t=5450, t=8000, t=100,000, t=1,000,000 and t=10,000,000.}
\end{figure}

\indent However, not every pattern generated by the Swift-Hohenberg equation results in the same type of hyperuniformity. By tuning the selected wavelength (or equivalently, by tuning $\epsilon$ in the original form Eq. (\ref{sh})), one can manipulate the ``persistence length" of the channels, which finally reveals the degree of disorder in these systems. To elucidate this point, we measure the scaling of scalar field fluctuations $\sigma_F^2(a)$ in these systems. We generate very large systems (10000$\times$10000) to perform the calculations. The comparison of four different systems are shown in Fig. 12. Notice that when channels get narrower, fluctuations decrease faster. The scaling ranges from the ordinary case ($a^{-2}$) to a hyperuniform case ($a^{-3}$).\\   
\indent The rapid drop of field fluctuations at small values of $a$  could be attributed to the fact that at short lengths scales, channels look rather ordered. Indeed, we model this behavior by introducing a toy ``polycrystal" model. We partition an entire system into many disjoint rectangular subsystems with a certain size distribution and then, inside each subsystem, put down parallel aligned channels with randomly chosen orientations. The behavior of the field fluctuations obtained from this model is similar to those seen in the Swift-Hohenberg patterns. Importantly, by applying this ``polycrystal" model as the initial condition, we find that the Swift-Hohenberg equation finally gives a solution with a higher peak in spectral densities (see Appendix B). This suggests that the solution of the Swift-Hohenberg equation is even more ordered than our toy ``polycrystal'' model. A possible explanation is that during the evolution the originally linear channels become tortuous so that they can connect themselves to the ones in adjacent subsystems that are formerly detached from each other, thus eliminate the defects at the boundaries in the toy ``polycrystal" model.\\

\begin{figure}[H]
\centering
\subfigure[]{
\includegraphics[width=7cm, height=4.7cm]{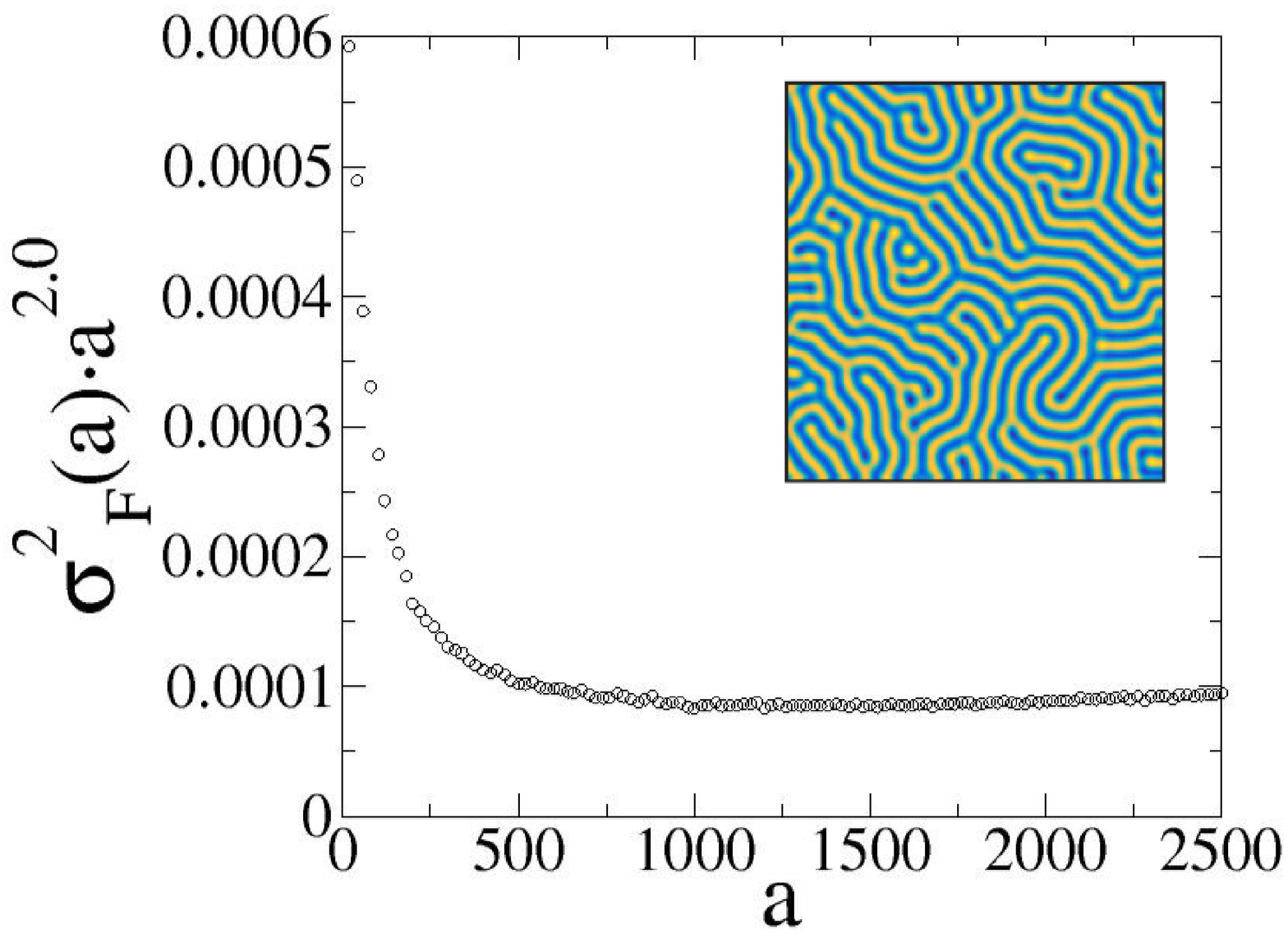}
}
\subfigure[]{
\includegraphics[width=7cm, height=4.7cm]{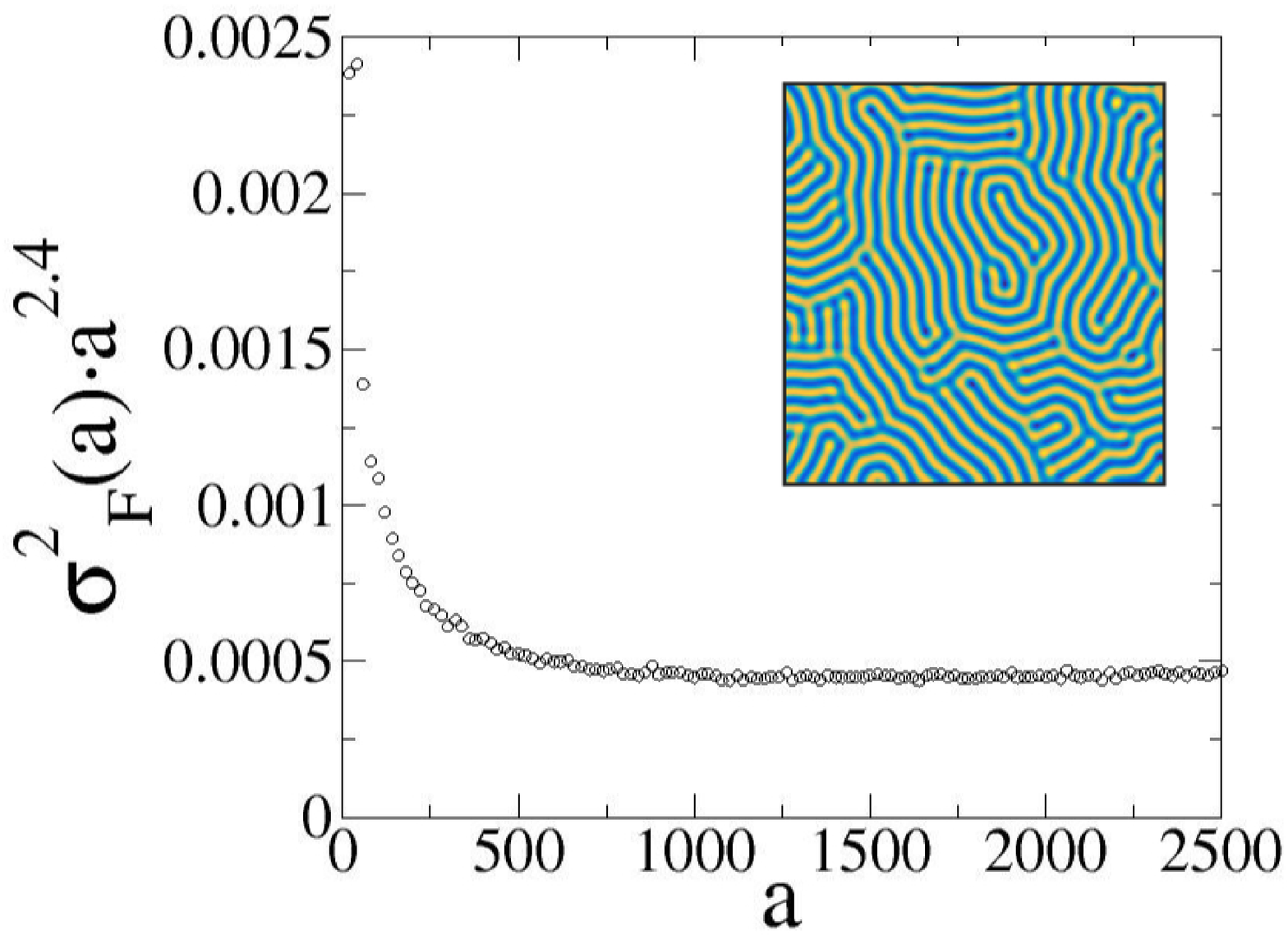}
}
\subfigure[]{
\includegraphics[width=7cm, height=4.7cm]{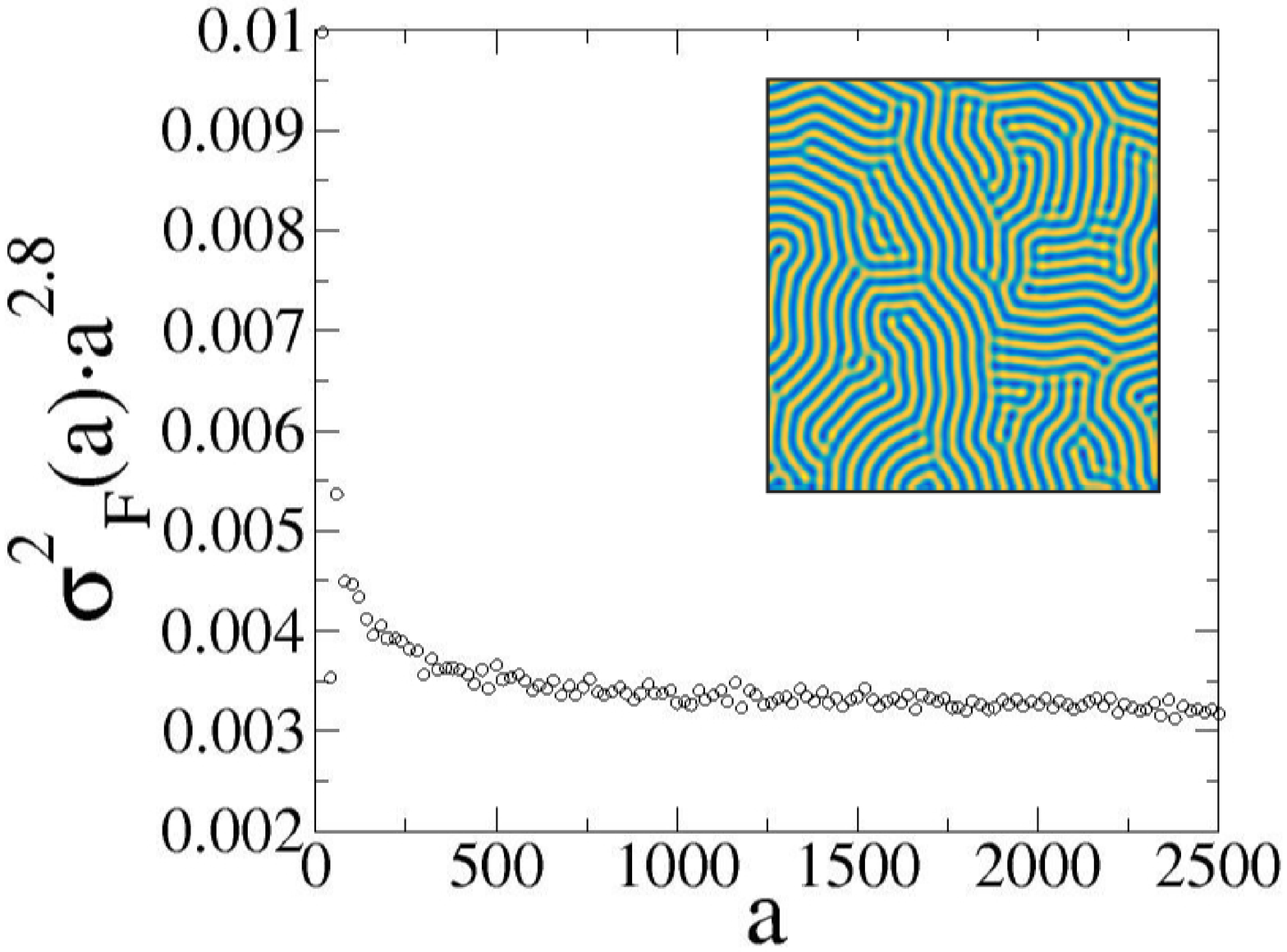}
}
\subfigure[]{
\includegraphics[width=7cm, height=4.7cm]{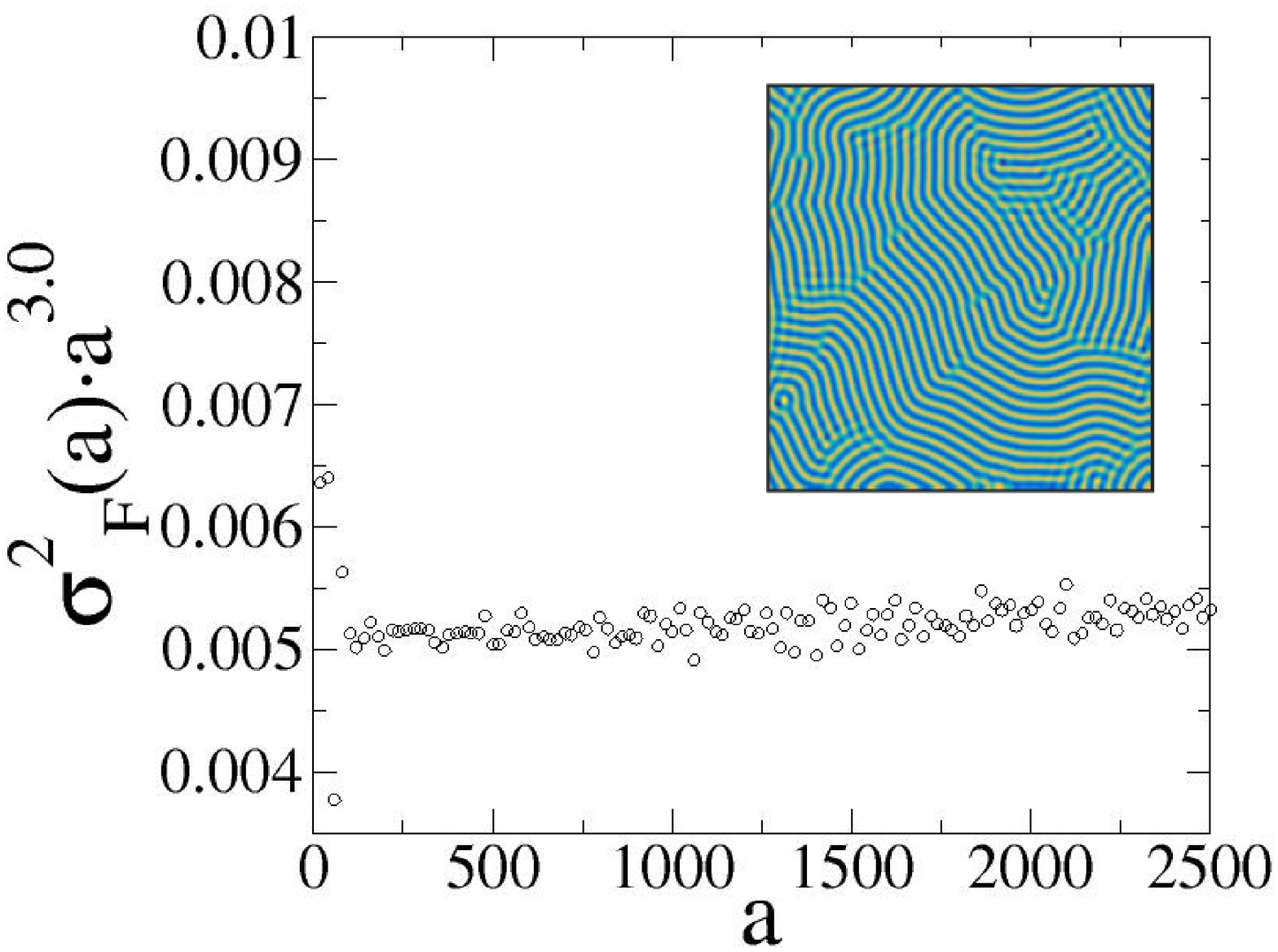}
}
\caption{(Color online.) Local scalar field fluctuations $\sigma_F^2(a)$ as a function of the window side length $a$ at t=100,000 with different $k_0$ as obtained from simulation of the Swift-Hohenberg equation. In each right-upper corner a very small portion (200$\times$200) of the system is shown. (a) Local field fluctuations with $k_0=0.2\pi$, the scaling is $a^{-2}$ for large-$a$. (b) Local field fluctuations with $k_0=0.7$, the scaling is $a^{-2.4}$ for large-$a$. (c) Local field fluctuations with $k_0=0.25\pi$, the scaling is $a^{-2.8}$ for large-$a$. (d) Local field fluctuations with $k_0=0.32\pi$, the scaling is $a^{-3}$ for large-$a$.}
\end{figure}

\section{LEVEL CUTS OF RANDOM SCALAR FIELDS}

\indent Our interest in this section is to determine whether two-phase random media derived from 
thresholding hyperuniform random scalar fields inherit the hyperuniformity property of the progenitor scalar fields. 

\subsection{$n$-Point Statistics from Gaussian random fields}

\indent We first consider the question for Gaussian random fields. For a Gaussian random field $F(\mathbf x)$, the joint probability density function is given by the following expression \cite{torquato2013random}
\begin{equation} \label{pn}
P_n[F(\mathbf x_1),...,F(\mathbf x_n)]=\frac {1}{\sqrt{{(2\pi)}^n\det \mathbf G}} \exp{(-\frac{1}{2}\mathbf F^T\mathbf G^{-1}\mathbf F)},
\end{equation} 
where elements of the correlation matrix $\mathbf G$ are given by the field-field correlation function
\begin{equation} \label{gij}
G_{ij}=G(x_{ij})=\left\langle F(\mathbf x_i)F(\mathbf x_j) \right\rangle,
\end{equation}
where $x_{ij}=|\mathbf x_j-\mathbf x_i|$. Without loss of generality, we consider Gaussian random fields with mean zero and variance unity. Then the field-field correlation function is actually the autocovariance function $\psi(\mathbf r)$, which is defined by Eq. (\ref{auto}). For statistically homogeneous fields, it is easily seen that all of the higher-order statistics of the field are determined by two-point information, namely, $\psi(\mathbf r)$. Furthermore, we restrict ourselves to isotropic scalar fields, implying that the autocovariance function $\psi(\mathbf r)$ is simply a radial function $\psi(r)$, where $r\equiv |{\bf r}|$.\\
\indent Now suppose we set a threshold $F_0$ to make the field a two-phase medium, $F_0$ can be taken as any real value between $-\infty$ and $\infty$. For phase 1 with $F>F_0$, one can write down the $n$-point probability function $S_n$ \cite{torquato2013random} 
\begin{widetext}
\begin{equation} \label{sn2}
S_n(\mathbf x_1,\mathbf x_2,...,\mathbf x_n)=\int_{-\infty}^{\infty}\cdotp \cdotp \cdotp \int_{-\infty}^{\infty}\{ \prod_{i=1}^n \Theta[F(\mathbf x_i)-F_0] \}P_n[F(\mathbf x_1),...,F(\mathbf x_n)]dF_1\cdotp \cdotp \cdotp dF_n.
\end{equation}
\end{widetext}
\indent \indent We are particularly interested in the first two lowest-order $n$-point probability functions, which enables us to obtain the autocovariance function $\chi_{_V}(r)$. It can be shown that \cite{torquato2013random, berk1991scattering}
\begin{equation} \label{chiv2}
\chi_{_V}(r)=S_2-S_1^2=\frac {1}{2\pi}\int_{0}^{\psi(r)}\exp(-\frac {F_0^2}{1+t})\frac {dt} {\sqrt{1-t^2}},
\end{equation}
which cannot be evaluated analytically in a closed form, except for $F_0=0$, which gives $\chi_{_V}(r)=\arcsin(\psi(r))/(2\pi)$. Without loss of generality, we expand the integral as a series expansion in the variable $\psi$ to analyze its properties, i.e.,
\begin{equation} \label{chivexp}
\chi_{_V}(r)=\frac{1}{2\pi}\exp(-F_0^2)\sum\limits_{n=1}^{\infty}a_n{\psi}^n,
\end{equation}
we find that the $n$th coefficient $a_n$ is
\begin{equation} \label{an}
a_n=\frac {1}{n!} {H_{n-1}}^2(F_0),
\end{equation}
where $H_{n-1}$ is Hermite polynomial of degree $n-1$. The lowest a few coefficients are $a_1=1$,
$a_2=F_0^2/{2!}$, $a_3=(F_0^2-1)^2/{3!}$ ...\\
\indent We now show that two-phase random media derived from a large class of hyperuniform Gaussian random fields that cannot be hyperuniform. Before formalizing this result as a proposition, we need to introduce some definitions and restrictions. In analogy with disordered sphere packings, \cite{torquato2006new} we define a statistically homogeneous and isotropic scalar field to be disordered if it possesses a radial autocovariance function $\psi(r)$ that decays to zero faster than $1/|{r}|^d$ for large $|{r}|$. This implies that the scalar field cannot have long-range order and that the autocovariance function is absolutely (Lebesgue) integrable.\\

{\it Proposition: A two-phase random medium derived from a hyperuniform disordered Gaussian random field cannot be hyperuniform.}\\\\
\indent Sketch of Proof: First, we observe that since $-1\leq \psi(r)\leq1$, for any integer $n$,
\begin{equation} \label{moments}
|\psi^n(r)| \leq |\psi(r)| \quad \mbox{for all}\quad r.
\end{equation}
Indeed, except at the zeros of $\psi(r)$, denoted by $\{z_1, z_2, \ldots\}$, and the origin, one has the following strict inequality for any $n \ge 2$:
\begin{equation} \label{moments2}
|\psi^n(r)| < |\psi(r)| \quad \mbox{for all}\quad r \notin \{0,z_1,z_2,\ldots\}.
\end{equation}
Moreover, because $\psi(r)$ is a decaying function, $\psi^n(r)$ is generally a different function from $\psi(r)$ that is skewed to the left for $n\geq2$ such that it satisfies (\ref{moments}). Let us rewrite the series expansion as 
\begin{equation} \label{expan}
\chi_{_V}(r)=\frac{\exp(-F_0^2)}{2\pi}[\psi(r)+\Delta \psi(r)],
\end{equation}
where 
\begin{equation} \label{delta}
\Delta \psi(r)=a_2\psi^2+a_3\psi^3+a_4\psi^4+\cdot\cdot\cdot
\end{equation}         
Assuming that the Gaussian random field is hyperuniform, that is $\int_{\mathbb R^d}{\psi(r)}d\mathbf r=0$, Eq. (\ref{expan}) yields the integral relation
\begin{equation} \label{positive}
\int_{\mathbb R^d} \chi_{_V}(r)d\mathbf r=\frac{\exp(-F_0^2)}{2\pi}\int_{\mathbb R^d}{\Delta \psi(r)}d\mathbf r\geq0.
\end{equation}
The nonnegativity of the integrals in (\ref{positive}) follows from the fact that the first integral is the spectral density of a two-phase medium at the origin. Since $\psi^n(r)$ is generally a different function from $\psi(r)$ that is skewed to the left for any $n\geq2$ such that the inequality (\ref{moments}) is obeyed, and each distinct power of $\psi(r)$ is different from any other power of $\psi(r)$, the volume integrals of $\psi^2(r)$, $\psi^3(r)$, $\psi^4(r)$, ..., will generally be positive, and hence 
\begin{equation} \label{proof}
\int_{\mathbb R^d} \chi_{_V}(r)d\mathbf r>0,
\end{equation}
i.e., the two-phase medium is not hyperuniform.\\ \\
{\it Remarks:}

1. Note that  $F_0=0$ represents the easiest case to allow for the possibility of hyperuniform two-phase media because even powers of the autocovariance (e.g., $\psi^{2m}(r)$ for $m \ge 2$),
which are intrinsically positive, do not appear.

2. In Appendix C, we provide explicit examples of realizable hyperuniform Gaussian autocovariance
functions and volume integrals of some of their high-order powers.

\subsection{Numerical results for Gaussian random fields} 

\indent The general argument above shows that a thresholded Gaussian random field cannot yield a hyperuniform two-phase system, even when the scalar field itself is hyperuniform. Indeed, this is consistent with numerical
results obtained here by thresholding the two hyperuniform scalar fields in Figs. 2 and 3. We found that after thresholding at zero, the hyperuniformity was destroyed for both cases. Two-phase configurations for both cases are shown in Fig. 13. The analysis reveals that the hyperuniformity metric $H$ increases by three orders of magnitude after thresholding, which means that the hyperuniformity is significantly destroyed by thresholding.  
\begin{figure}[H]
\centering
\subfigure[]{
\includegraphics[width=3.8cm]{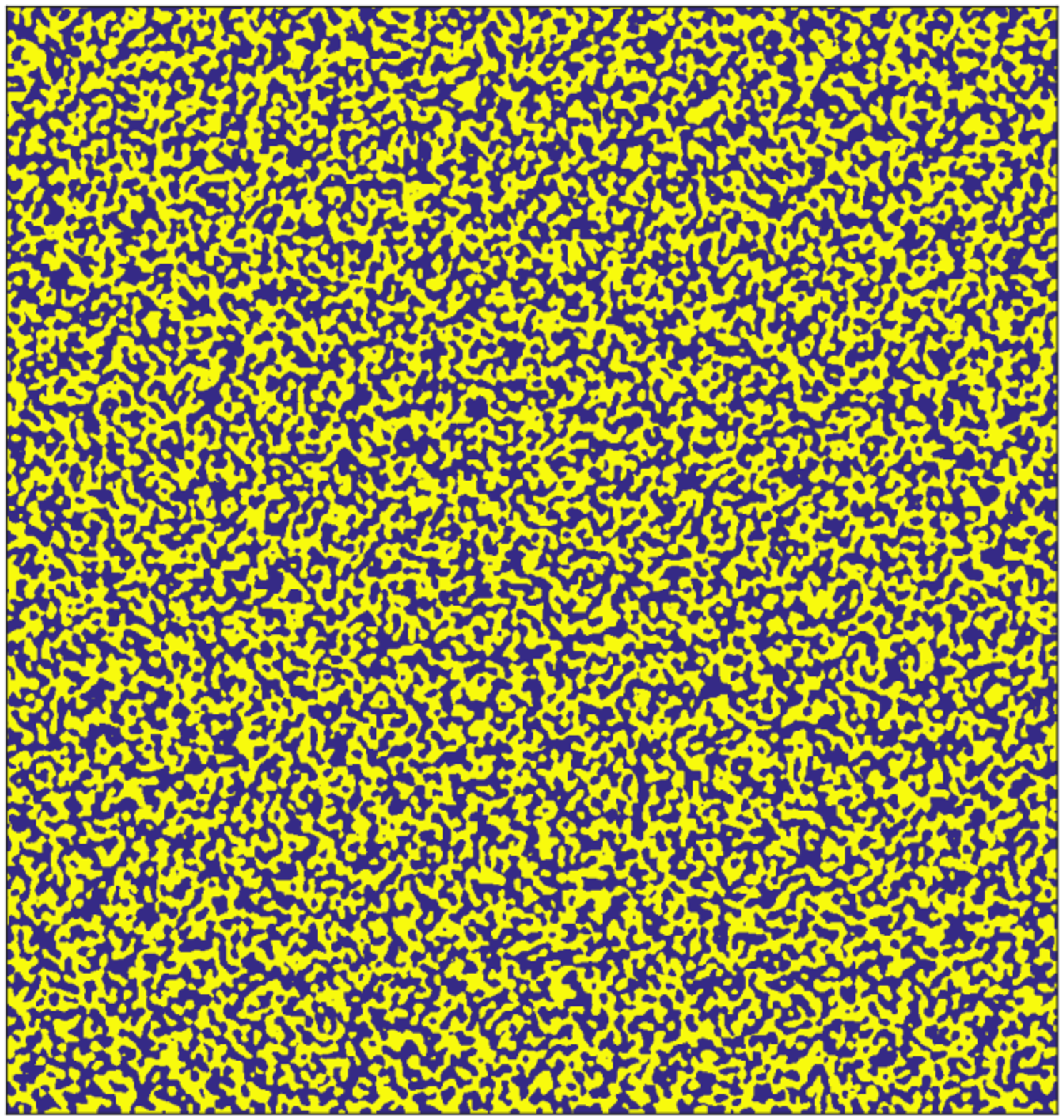}
}
\subfigure[]{
\includegraphics[width=3.8cm]{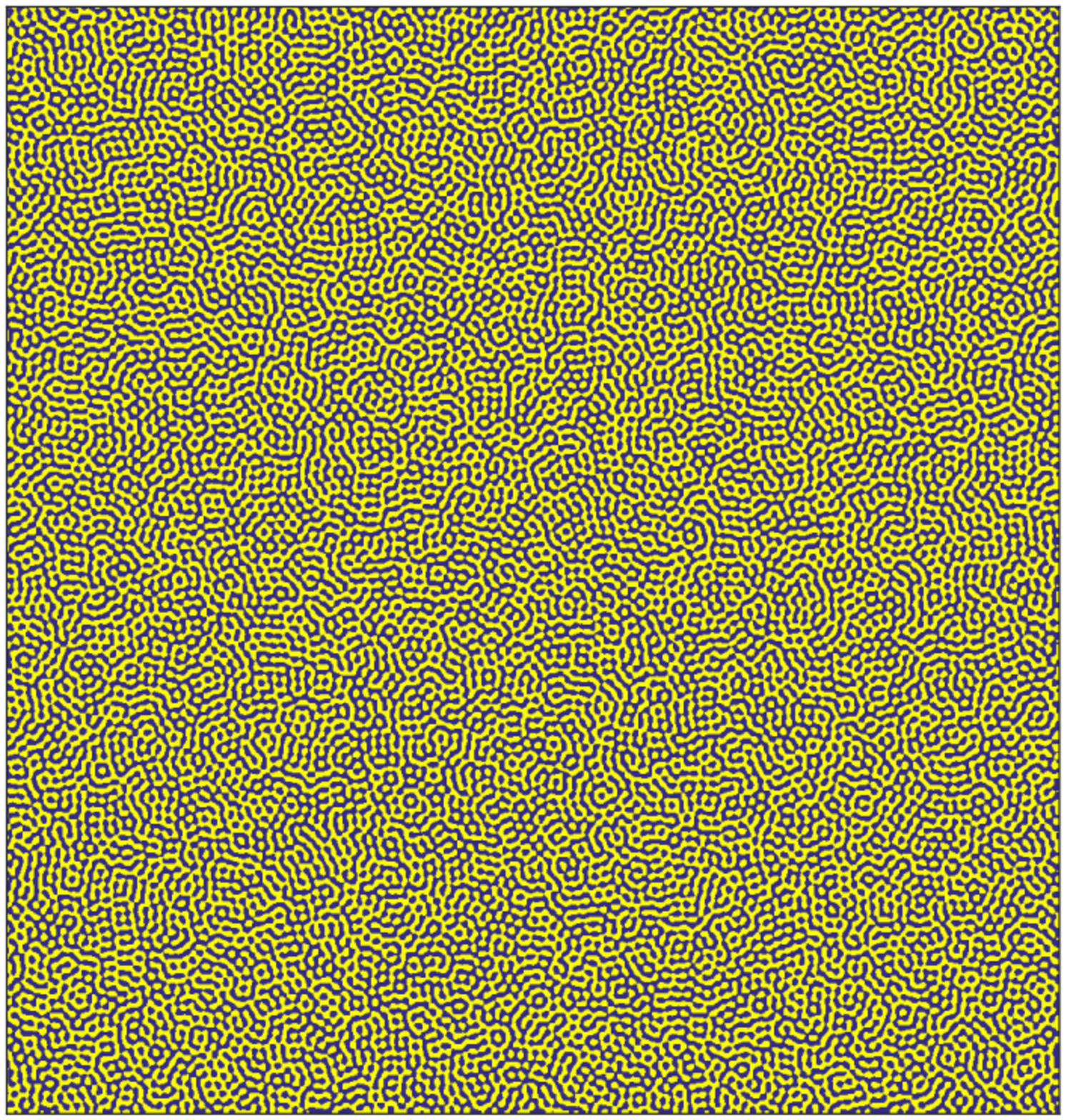}
}
\caption{(Color online.) Thresholded two-phase configurations from Sec. III. The threshold is zero, and yellow phase is positive while the blue phase is negative. The system size is $1000\times1000$ for both cases. (a) Thresholded two-phase configuration obtained from the Gaussian field shown in Fig. 2. The extrapolated spectral density at the origin is 0.47, which gives the hyperuniformity metric $H \approx 0.06$. (b) Thresholded two-phase configuration obtained from the Gaussian field shown in Fig. 3. The extrapolated spectral density at the origin is 0.25, which gives the hyperuniformity metric $H \approx 0.01$.}
\end{figure}

\subsection{Towards hyperuniform two-phase media from non-Gaussian fields}

\indent The practical question remains as to how close one can come to achieve perfect hyperuniformity? Of course, a thresholded periodic scalar field is still hyperuniform. However, a periodic scalar field is neither isotropic above dimension one, nor disordered according to our aforementioned criterion. We expect that a hyperuniform non-Gaussian scalar field that is essentially a two-phase system (bimodal distribution with sharp peaks) should be more likely to be hyperuniform after thresholding. Examples of such patterns are spinodal decomposition structures introduced and studied in Sec. III and IV. In Fig. 14, we show a comparison of the original spectral density (the whole profile of the one depicted in Fig. 9) and the corresponding one after thresholding. The functional forms of the spectral densities are nearly the same (up to a vertical-axes scaling factor), as expected. The only difference is that the low-$k$ scaling is lost after thresholding. Even so, the low-$k$ values of the spectral densities still remains very small compared to the peak value, giving the hyperuniformity metric $H \approx 10^{-3}$. 
We expect that the discrepancy at low $k$ would be made smaller  by employing a higher resolution in the thresholded system.\\ 
\indent Moreover, as noted in the Introduction, we know that scalar fields generated by convolving hyperuniform point configurations with a non-negative dimensionless radial scalar kernel function $K(\mathbf x;\mathbf C)$, where $\mathbf C$ represents a set of parameters that characterizes the shape of the radial function, will inherit the hyperuniformity of the original point configurations. \cite{PhysRevE.94.022122} Scalar fields generated in this way are generally non-Gaussian. It is clear that in the case in which the kernel function is highly concentrated at each of the points in the point patterns, the resulting non-Gaussian scalar field is nearly a two-phase system. After thresholding such a system, the resulting two-phase medium will consists of ``particles" distributed throughout a ``matrix" that inherits the hyperuniformity of the original point configuration. This provides another possible method to obtain disordered hyperuniform two-phase media from non-Gaussian random scalar fields.\\ 
\begin{figure}[H]
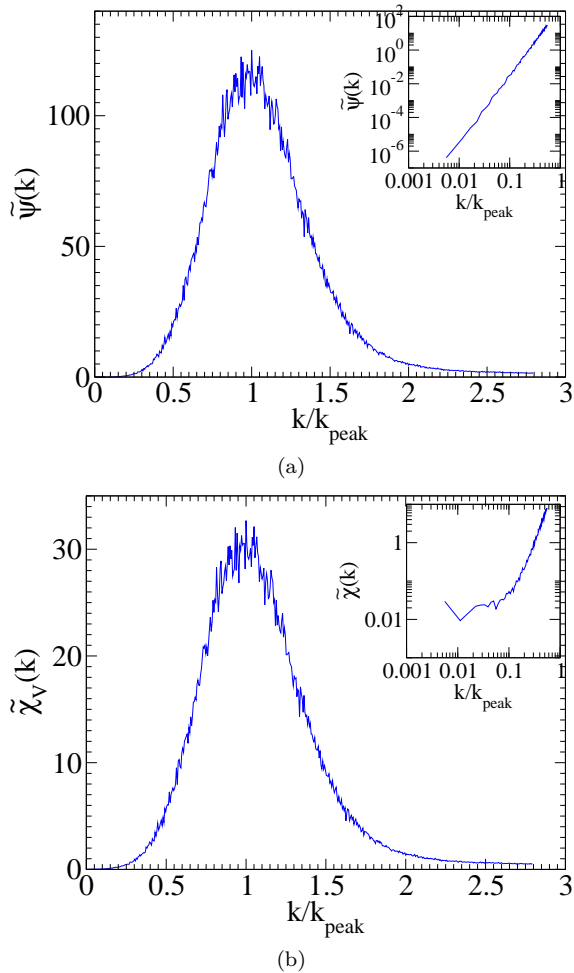

\centering
\subfigure[]{
\includegraphics[clip,width=7.5cm]{fig14a.eps}
}
\subfigure[]{
\includegraphics[clip,width=7.5cm]{fig14b.eps}
}
\caption{(Color online.) A comparison of spectral densities before and after thresholding at zero. The underlying configuration was obtained from the same 10000$\times$10000 system at t=100,000 from the Cahn-Hilliard equation that gives the results shown in Figs. 8 and 9. The insets show the behavior near the origin. (a) Angular averaged spectral density of the scalar field. (b) Angular averaged spectral density of the thresholded two-phase system.}
\end{figure}

\section{CONCLUSIONS AND DISCUSSION}

\indent The hyperuniformity concept has been generalized to theoretically characterize two-phase media, scalar fields, and random vector fields. \cite{0953-8984-28-41-414012, PhysRevE.94.022122} In this paper, we focused on the study of hyperuniform scalar fields, especially how to construct realizations of such systems via Gaussian random fields as well as spatial patterns that emerge as solutions to the Cahn-Hilliard equation for spinodal decomposition, and Swift-Hohenberg equations. We re-examined these classical models under the ``hyperuniform" lens. Our results advance our theoretical understanding of
hyperuniformity and provides new guidelines to synthesize these new classes of hyperuniform materials. We showed that one can construct hyperuniform (or stealthy) Gaussian fields by directly designing their spectral densities and our approximated numerical examples demonstrated that they are indeed effectively hyperuniform. \\
\indent We also analyzed the time evolution of spinodal decomposition patterns as obtained from the Cahn-Hilliard equation. We showed that such systems are hyperuniform after they enter the scaling regime, but hyperuniformity may emerge even earlier. We look forward to seeing further experimental work that support our findings, especially for the case of off critical quenches. It is expected that experimentalists can now produce hyperuniform materials directly by ``freezing" the dynamics. Since the Cahn-Hilliard equation is widely applied in physics, chemistry, biology \cite{Saranathanrsif20120191} and even ecology, \cite{liu2013phase} it also implies that hyperuniformity may be more universal than has been realized thus far. Indeed, many patterns in ecology have been well modeled by the Turing-like reaction-diffusion mechanism mentioned earlier. \cite{59776} Remarkably, a recent paper that studied the interplay between termite mounts and vegetation reports a spectral density that appears to indicate vegetation patterns between termite mounds are hyperuniform, \cite{bonachela2015termite} but this was not central to the paper and requires further study.\\ 
\indent We also studied pattern formation that are derived from solutions of the Swift-Hohenberg equation. We demonstrated that such patterns can indeed be hyperuniform. We found that certain labyrinth-like patterns that emerge as solutions to the Swift-Hohenberg equation are effectively hyperuniform and actually exhibit spectral densities that resemble stealthy ones.\\
\indent We also studied the hyperuniformity of two-phase media derived from level cuts of hyperuniform scalar fields. We showed that a thresholded Gaussian random field cannot be hyperuniform in general. However, nearly hyperuniform two-phase media may be obtained from other systems. Results of spinodal decomposition patterns were presented to compare with the ones from Gaussian random fields. We showed that thresholded hyperuniform spinodal decomposition spatial patterns are near effectively hyperuniform. This implies that to generate a hyperuniform two-phase system from a hyperuniform scalar field, it is advantageous for the latter to be a bimodal field with relatively sharp peaks. A thorough study of level cuts of scalar fields is expected to enable experimentalists to use state-of-the-art techniques, such as stereolithography and 3D printing, to synthesize hyperuniform materials with novel physical properties.\\
\indent Our study shows that hyperuniformity can emerge in scalar fields. The quantification of long-wavelength scalar field fluctuations 
and the hyperuniformity metric $H$ 
provide useful ways to characterize the degree of global order of scalar fields and hence enables the classification of wide class of spatial patterns. This work paves the way 
for investigators to revisit many patterns found across different disciplines (e.g, physics, chemistry, biology and ecology) 
and reexamine them under the hyperuniformity lens. It is expected that such future work should enable one to gain new insights 
about the general mechanisms that lead to hyperuniformity as well as expand experimental 
capabilities to make large samples of new classes of hyperuniform materials.\\

\begin{acknowledgements}
\indent The authors are grateful to Paul Chaikin, Duyu Chen, Michael Cross, Pierre Hohenberg, Michael Klatt and James Stone for helpful discussions. This work was supported in part by the Materials Research Science and Engineering Center (MRSEC) program of the National Science Foundation under Award Number DMR-1420073.\\
\end{acknowledgements}
                   
\section{APPENDIX}

\subsection{Generalized Cahn-Hilliard Equations and Hyperuniformity}

\indent Here we consider generalized Cahn-Hilliard-type equations and propose some open questions with regard to generating hyperuniform random scalar fields.\\
\indent In Sec. III, the free energy density of the Cahn-Hilliard equation can be defined as a functional, i.e.,
\begin{equation} \label{freeenergy}
\rho(c)=\frac {1} {4} (c^2-1)^2+\frac {\gamma}{2}{|\nabla c|}^2.
\end{equation}
\indent It is clear that $c=\pm1$ are two energy minima of $\frac {1} {4} (c^2-1)^2$. Physically the form of this functional can be seen as the direct result of the Ginzburg-Landau theory with certain symmetries. However, we can use other forms of $f(c)$, which gives the free energy density
\begin{equation} \label{freeenergy2}
\rho(c)=f(c)+\frac {\gamma}{2}{|\nabla c|}^2.
\end{equation}
\indent And the time evolution equation will be
\begin{equation} \label{newch}
\frac{\partial{c}}{\partial{t}}=D\nabla^2(f^\prime(c)-\gamma\nabla^2c).
\end{equation}

\begin{figure}[H]
\centering
\subfigure[]{
\includegraphics[width=5.5cm, height=5.5cm]{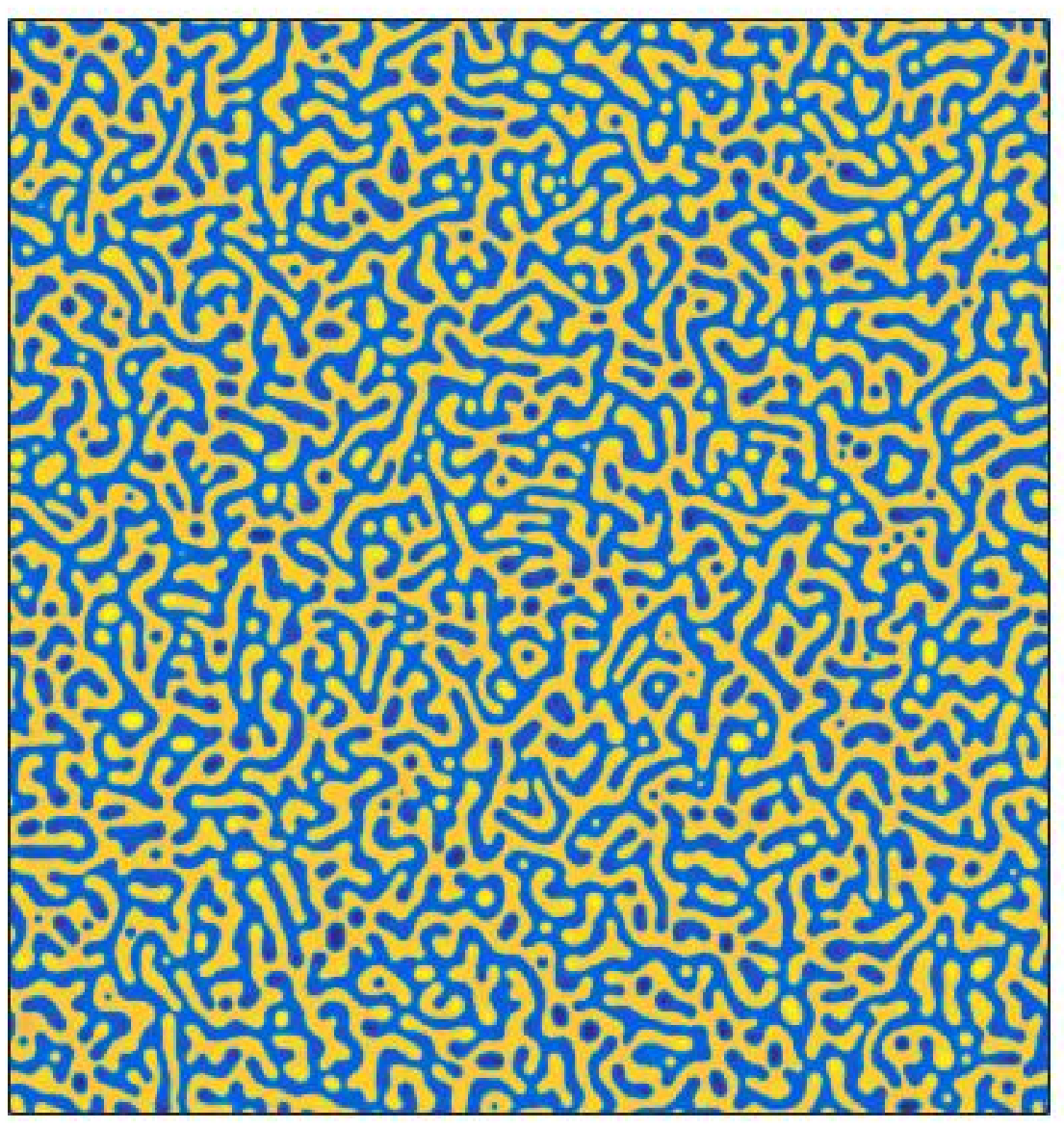}
}
\subfigure[]{
\includegraphics[width=6.5cm, height=4.5cm]{fig15b.eps}
}
\caption{(Color online.) (a) The scalar field at t=100,000 from the time evolution of Eq. (\ref{newch}). (b) The corresponding angular averaged spectral density of (a).}
\end{figure}

\indent As a simple example, we can use $f(c)=\frac {1}{8} (c^2-1)^4$, which retains two minima as the original one. Follow the same set up, we can observe the time evolution of the system started from the random initial condition. The result is shown in Fig. 15. The hyperuniformity metric $H$ is larger than $10^{-3}$, but the spectral density does have a trend that goes to zero. One can easily come up with many different forms of $f(c)$, and it is of interest to see if the dynamics will give rise to hyperuniform or even stealthy random scalar fields.

\subsection{Toy ``Polycrystal" Model}

\indent Some detailed results of the toy ``polycrystal" model introduced in Sec. IV are presented here. For illustration purposes, we consider here a $1000\times1000$ ``polycrystal" consists of 100 subsystems, the subsystems are of equal size $100\times100$. The pixels are assigned a value of 1 or 0 determined by which phase the pixels are in. The width of the stripe is 5 pixels and then the characteristic wavelength is 10 pixels. We compare the spectral density of this system with the one directed generated by the Swift-Hohenberg equation at $t=100,000$ with the same system size and characteristic wavelength, which means that $k_0$ is chosen to be $2\pi/10$. In order to compare patterns generated by the Swift-Hohenberg equation with the toy ``polycrystal" model, all patterns are thresholded at the value of 0 to convert to two-phase systems. We also use the polycrystal configuration as our initial condition to see how it evolves under the Swift-Hohenberg equation. We stop these process at $t=1,000,000$ and find that the originally linear stripes become distorted and the defects at the boundaries between different subsystems disappear, and stripes in adjacent subsystems connect with each other at the boundaries (see Fig. 16). We show the comparison of the spectral densities of the three systems mentioned above in Fig. 17. Note that the spectral densities of our toy model resemble the ones of the Swift-Hohenberg equation, suggesting that the model is a good description of generated patterns. The higher peak of the pattern modified by the Swift-Hohenberg equation demonstrates that the dynamics helps reduce the defects at the boundaries and as a consequence the system becomes more ordered.        
\begin{figure}[H]
\centering
\includegraphics[width=4.5cm]{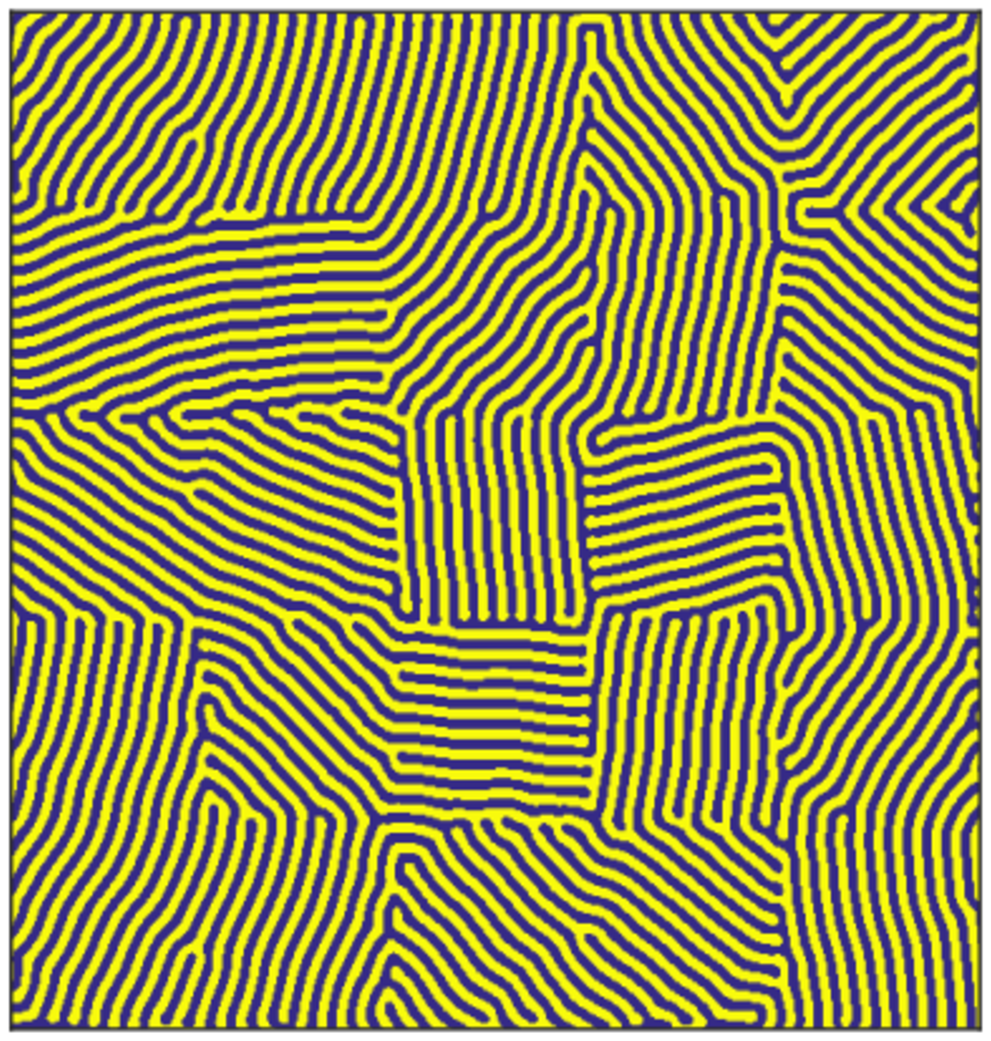}
\caption{(Color online.) The system under the evolution of the Swift-Hohenberg equation with $k_0=2\pi/10$ at t=1,000,000 using the ``polycrystal" initial condition. A $500\times500$ portion is shown here.}
\end{figure}

\begin{figure}[H]
\centering
\includegraphics[width=6cm, height=4.5cm]{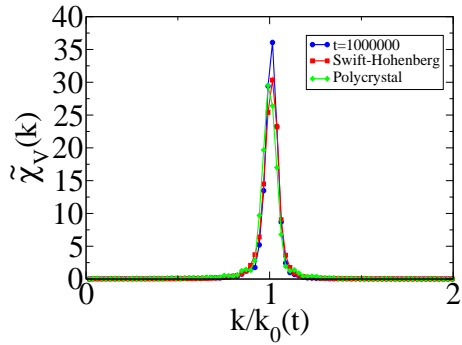}
\caption{(Color online.) The comparison of spectral densities of three two-phase systems. The green one is of the toy ``polycrystal" model. The red one is from thresholding a typical pattern generated by the Swift-Hohenberg equation with $k_0=2\pi/10$ at $t=100,000$. The blue one is from thresholding the pattern under the evolution of the Swift-Hohenberg equation of $k_0=2\pi/10$ at $t=1,000,000$ using the ``polycrystal" initial condition. The thresholds for the later two cases are both zero.}
\end{figure}

\subsection{Examples of Autocovariance Functions and Volume Integrals of Their Powers}

\indent In Sec. V.A, we proved that the hyperuniformity sum rule for two-phase media (\ref{sum2p}) derived from thresholding hyperuniform Gaussian random fields generally does not hold. We show this by expanding $\chi_{_V}(r)$ as a series of $\psi(r)$ and show that the volume integral of each term generally gives positive contributions because the higher-order powers of $\psi$ are shifted to the
left such that they satisfy the inequality (\ref{moments2}).\\
\indent Here we illustrate these properties by considering two explicit realizable hyperuniform Gaussian autocovariance function in three dimensions, one of which is a damped-oscillating function:
\begin{equation} \label{ex1}
\psi_1(r)=\sqrt{2}\exp(-r)\cos(r-\frac {\pi}{4}),
\end{equation}
and the other a function with a single zero that decays to zero with the power law $1/r^4$:
\begin{equation} \label{ex2}
\psi_2(r)=\frac {4}{3(r+1)^5}-\frac{1}{3(r+1)^4},
\end{equation} 
In Fig. (18), we show the plots of both functions and their 2nd to 5th powers.\\

\begin{figure}[H]
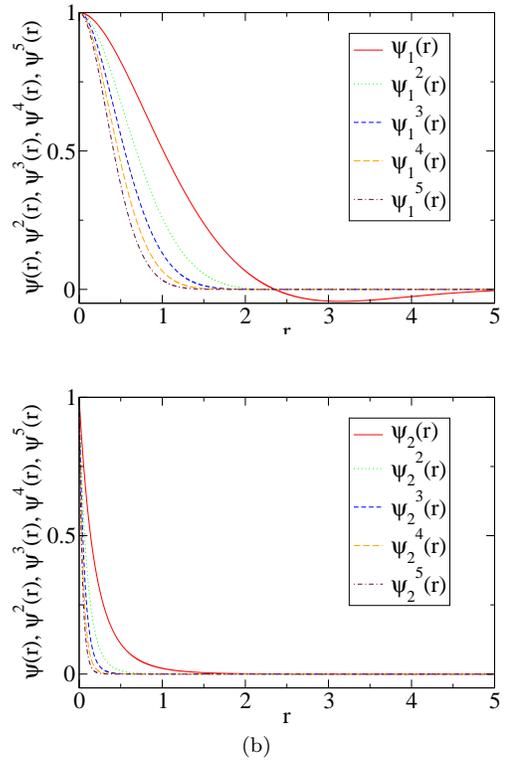

\centering
\subfigure[]{
\includegraphics[width=6.5cm]{fig18a.eps}
}
\subfigure[]{
\includegraphics[width=6.5cm]{fig18b.eps}
}
\caption{(Color online.) (a) Autocovariance function $\psi_1(r)=\sqrt{2}\exp(-r)\cos(r-\pi/4)$ and the square, the cube, the 4th power and the 5th power of $\psi_1(r)$. (b) Autocovariance function $\psi_2(r)=(4/(r+1)^5-1/(r+1)^4)/3$ and the square, the cube, the 4th power and the 5th power of $\psi_2(r)$.}
\end{figure}

\indent Although these autocovariance functions have distinctly different properties, it is clear that these higher powers are skewed to the left and the net positive contribution of the volume integral is larger than that of the volume integral of $\psi(r)$, which is identically zero. As expected, the volume integrals of these higher powers are all positive; specifically, $\int \psi_1^2(r) d{\mathbf r}=5\pi/4$, $\int \psi_1^3(r) d{\mathbf r}=2032\pi/3375$, $\int \psi_1^4(r) d{\mathbf r}=3033\pi/8000$, $\int \psi_1^5(r) d{\mathbf r}=2831200\pi/10793861$, $\int \psi_2^2(r) d{\mathbf r}=32\pi/2835$, $\int \psi_2^3(r) d{\mathbf r}=3392\pi/1216215$, $\int \psi_2^4(r) d{\mathbf r}=8458\pi/695020095$ and $\int \psi_2^5(r) d{\mathbf r}=59264\pi/113730561$.\\

\end{document}